# High-Energy Atmospheric Radiation: From Thunderstorm Ground Enhancements to Terrestrial Gamma-Ray Flashes

**A. Chilingarian, L. Hovhannisyan, B. Sargsyan, and M. Zazyan**

**Yerevan Physics Institute, Alikhanyan Brothers 2, Yerevan, Armenia, AM0036**

## Abstract

It is widely recognized that the atmosphere shields the Earth from harmful radiation; however, understanding its active role in particle acceleration is essential for research in space weather and atmospheric physics. Thunderstorms generate strong atmospheric electric fields (AEFs) that extend over large areas within and around storm systems. The energy in AEFs mainly converts into radiation energy through relativistic runaway electron avalanches (RREAs, Gurevich et al., 1992), which appear as thunderstorm ground enhancements (TGEs, Chingarian et al., 2010, 2011) when detected on Earth's surface and as gamma glows when observed above intense tropical thunderstorms. Variations of these primary radiation sources include short gamma ray bursts recorded by orbiting gamma observatories at 500-700 km altitude, called terrestrial gamma ray flashes (TGFs, Fishman et al., 1994), and downward terrestrial gamma ray flashes (DTGFs, Ortberg et al., 2024), recorded in coincidence with lightning leader propagation.

This work presents a unified conceptual and observational framework that reinterprets these radiation bursts as manifestations of the same runaway processes happening at different atmospheric depths (Dual-stage model, DSM). We review recent results from satellite (ASIM), aircraft (ALOFT), balloon (HELEN), and ground-based (SEVAN and KANAZAWA) experiments to demonstrate the advantages of this integrated approach. This study addresses key contradictions in the field, introduces new classification criteria based on physics rather than detector location, and enhances our understanding of particle acceleration in thunderstorms.

**Key Points:**

- **Unified physical mechanism: TGEs, gamma glows, TGFs, and DTGFs are all outcomes of the RREA process occurring in different atmospheric regions.**
- **Multi-platform data synthesis: Includes results from ASIM (space), ALOFT (aircraft), Kanazawa (DTGF), and SEVAN (TGE).**
- **A dual-stage, feed-forward coupling mechanism in which a lower-dipole RREA supplies photons that seed electrons to upward RREA (DSM model).**
- **A new model for DTGF origination: field reorganization, not leader-gap acceleration.**

- **Comparison of the measured electron TGE spectrum with theoretical expectations at different electric field strengths**

## 1. Introduction

Charge separation in thunderclouds, driven by updrafts of warm air and tension among hydrometeors, creates opposite dipoles within the cloud. C.T.R. Wilson, one of the first particle physicists and a pioneering researcher in atmospheric electricity, introduced a puzzling physical phenomenon known as "runaway" electrons at the start of the last century (see Chilingarian et al., 2025, and references therein). He proposed that free electrons are accelerated by the atmospheric electric field (AEF) into open space in the dipole between the main negatively charged region (MN) in the middle of the thundercloud and the main positive (MP) charge region at the top. However, Joachim Kuettner, conducting groundbreaking experiments at Zugspitze, discovered a more complex tripole charge structure within the charged layers of thunderclouds (Kuettner, 1950). According to the tripole model, the atmospheric electric field consists of upper and lower dipoles that accelerate free electrons both toward open space and the Earth's surface. Not everyone immediately understood the significance of Kuettner's discovery. In 1963, Richard Feynman wrote: "The top of the thunderstorm is positively charged and the bottom negative, except for a small local area of positive charge at the bottom of the cloud, which caused much concern for everyone. No one seems to know why he is there or how important he is. If it weren't for him, everything would be much easier" (Feynman et al., 1963). However, further investigations of cloud charge structure (Williams, 1989) confirm the tripole structure of cloud charge and the role of "graupel" hydrometeors in forming the transient LPCR. According to modern theories, LPCR is essential for charge separation within the cloud and the initiation of lightning (Nag & Rakov, 2009; Chilingarian et al., 2017a). Without it, lightning would most likely strike toward the upper positive charge. Thus, without this "useless" local area of positive charge, our planet might be quiet, dark, and lifeless.

Free electrons from Extensive air showers (EASs, Auger et al., 1939) are abundant in the troposphere. The altitude of the electron maximum density, known as the Regener–Pfotzer maximum, depends on the geomagnetic cutoff rigidity (Rc), and the phase and intensity of the solar cycle. Recent measurements, confirmed by PARMA4 calculations (Sato, 2016), show that the peak flux of charged particles occurs in equatorial regions (Rc = 10-15 GV) at an altitude of 16-17 km (Fig. 4 of Ambrozova et al., 2023). Electric fields generated by intense thunderstorms transfer energy to free electrons, accelerate them, and, under certain conditions, produce electron-photon avalanches. In 1992, Gurevich, Milikh, and Roussel-Dupré established conditions for the extensive multiplication of electrons from each energetic seed electron injected into the region of a strong electric field (Gurevich et al., 1992). This process is now called the Relativistic Runaway Electron Avalanche (RREA; Babich et al., 2001; Alexeenko et al., 2002). A numerical approach for solving the relativistic Boltzmann equation for runaway electron beams (Symbalisty et al., 1998) helps estimate the threshold (critical) electric field (Babich et al., 2001; Dwyer et al., 2003) required to initiate RREA. At standard temperature and pressure in dry air at sea level, Eth ≈ 2.80 ·n kV/cm (density-scaled threshold), where air density n is relative to the International Standard Atmosphere (ISA) sea-level value (see the recent update of the critical energy Eth ≈ 2.67·n kV/cm in Dwyer and Rassoul, 2024). This threshold field is slightly higher than the breakeven field, which corresponds to the electron energy at which minimum ionization occurs. If electrons traveled exactly along electric field lines, this would define the threshold for runaway electron propagation and avalanche formation. However, the paths of electrons deviate due to Coulomb scattering with atomic nuclei and Møller scattering with atomic electrons, leading to deviations from the ideal case. Moreover, secondary electrons produced by Møller scattering are not generated along the field line; therefore, electric fields 10-20% stronger are needed for electrons to run away and initiate an avalanche.In Figure 1, we illustrate the radioactive terrestrial atmosphere, which is bombarded by billions of cosmic rays (CR) entering from space and originating in the atmosphere and the Earth's crust. The atmospheric radiation sources are shown on the left side of the diagram. The lower dipole comprises the main negative layer and an induced mirror charge on the Earth's surface. Another dipole forms between the main negative layer and LPCR, which is often associated with falling graupel (snow pellets coated with a layer of ice, Kuettner, 1950). Electrons accelerated by the two lower dipoles generate electron-gamma ray avalanches, which are detected as thunderstorm ground enhancements (TGEs; Chilingarian et al., 2010, 2011). Due to the proximity of the electron accelerators in thunderclouds to the surface, TGEs contain millions of gamma rays, electrons, positrons, and neutrons. Particle detectors beneath electron accelerators at high-altitude sites in Armenia, Eastern Europe, Germany, and other locations have recorded hundreds of TGEs, confirming extensive particle fluxes over thousands of square kilometers and durations ranging from seconds to tens of minutes (Chum et al., 2020; Chilingarian et al., 2024a). Data from Aragats have convincingly validated the RREA/TGE model through direct measurements of electron and gamma-ray spectra, which have also detected atmospheric neutrons and positrons (Chilingarian et al., 2012, 2024b). Importantly, these measurements agree well with predicted energy spectra (Dwyer, 2003) and highlight the role of seed electrons from EASs.

The dipole between the LPCR and its mirror in the Earth accelerates positrons and positive muons while decelerating electrons and negative muons (Chilingarian et al., 2024a; 2024b). The LPCR is temporary and vanishes with the fall of graupel, causing rapid changes in the cloud's charge structure. This, in turn, quickly alters the AEF and the acceleration and deceleration modes of charged particles.

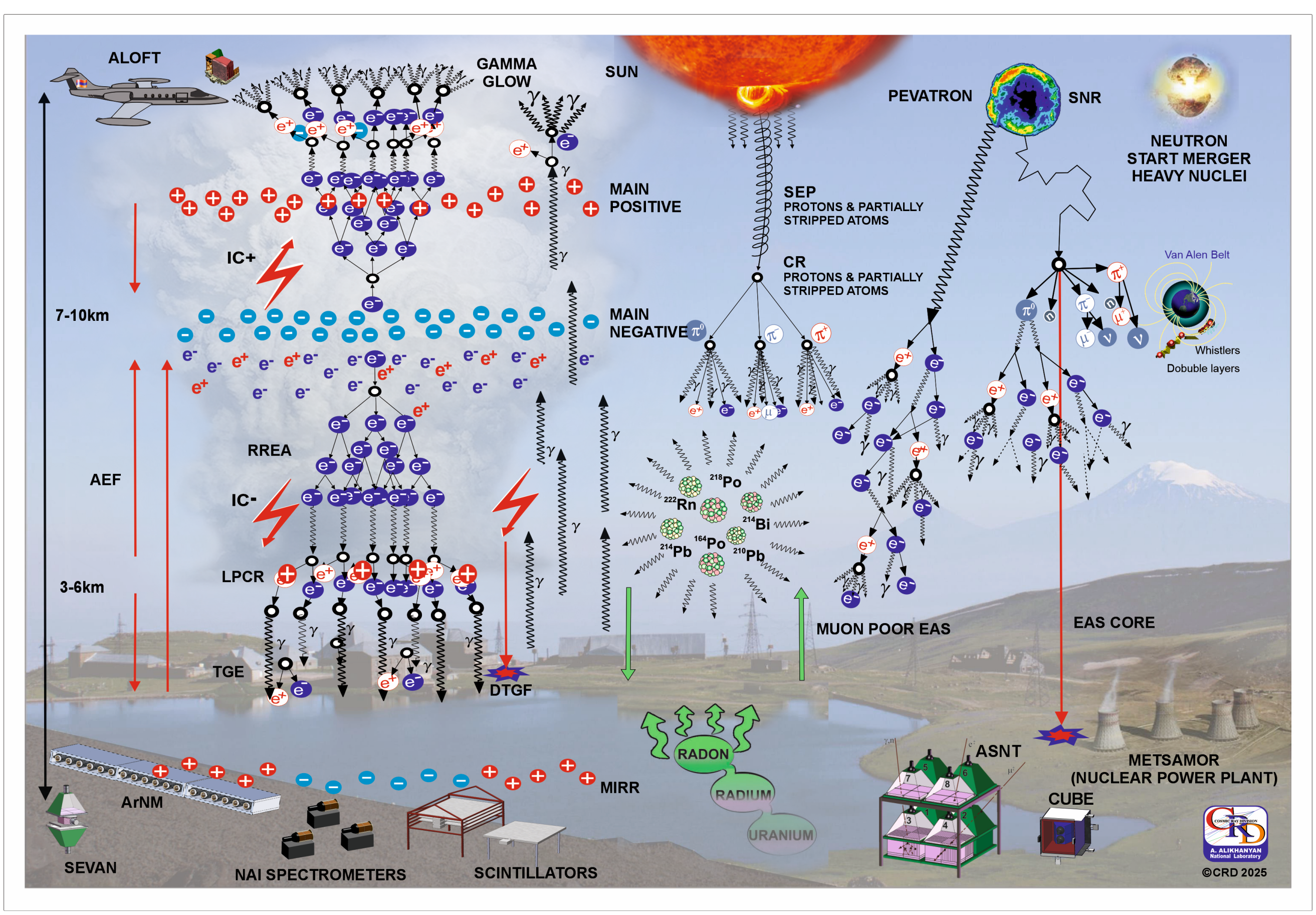

**Figure 1. The fluxes of secondary particles from space and atmospheric accelerators, as well as gamma radiation from 222Rn progeny. Additionally, the diagram shows the sources of primary cosmic rays. In the background, the Aragats cosmic ray station, situated at an elevation of 3200 meters, is equipped with various particle detectors and spectrometers.**

The upper dipole consists of the main negative and main positive charge layers (see Fig. 1). Electrons accelerated within this dipole generate avalanches in the upper atmosphere, leading to gamma-ray glows detected by airborne experiments flying above thunderstorms (Kelly et al., 2015; Ostgaard et al., 2019; Marisaldi et al., 2024; Helmerich et al., 2024). Additionally, some gamma rays reach orbiting gamma observatories 400-700 km away from the source, triggering microsecond bursts of particles known as terrestrial gamma flashes (TGFs, Fishman et al., 1994; Mailyan et al., 2015). The dipoles in the lower atmosphere rely on a relatively stable downward flux of free electrons from EASs as seeds to generate RREAs, which are observed on Earth's surface as TGEs. The difficulty in providing seeds to trigger RREA in the upper dipole has persisted for the past 30 years. It is not possible to reverse the downward electrons to initiate RREA in the upper dipole. Several proposed sources of seeds moving upward face significant difficulties (see review Chilingarian, 2024).

To overcome these challenges, we propose a two-stage, vertically coupled RREA mechanism that links a lower, downward-accelerating region—where strong RREAs generate bremsstrahlung—to an upward-accelerating region that continuously transforms gamma seed flux into gamma-ray glows observed in the upper atmosphere. We will demonstrate that a fraction of gamma rays from the lower dipole, redirected by Compton scattering, combined with conversion near the start of the upper dipole and kilometer-scale upper-dipole fields, can sufficiently account for the magnitude, duration, and altitude of upward glows without requiring a single, unrealistically bright gamma-ray source, usually introduced to explain TGF fluxes. (Mailyan et al., 2015)

The right side of Fig. 1 illustrates the EASs formed high in the atmosphere by galactic gamma rays, protons, and fully stripped nuclei interacting with atmospheric atoms. High-energy particles are produced during violent explosions and mergers within the galaxy and beyond. The diagram shows a supernova remnant and a neutron star merger. When cosmic-ray–induced extensive air showers develop in the atmosphere, their lateral spread can reach several hundred meters to several kilometers, depending on the primary energy and altitude. The cores of EAS contain the highest-energy secondary particles, producing very short, compact bursts, as indicated by the red circle. Muon-poor events detected at altitudes above 4000 meters suggest Pevatrons- stellar sources that accelerate protons up to $10^{15}$ eV. The illustration also features a nuclear power plant, a potential source of radioactive contamination, and the Van Allen belt, which can emit MeV electrons toward Earth.

## 2. Corsika simulations of RREAs reaching Aragats stations and comparisons with the electron flux recovered from observed TGEs

The direct evidence of the RREA process in a thunderous atmosphere is the detection of RREA electron flux. The first measurement of electron flux during thunderstorms at Aragats on September 19, 2009 (Chilingarian et al., 2017b), marked the beginning of High-Energy Physics in the Atmosphere (HEPA) research, which continues to this day. The historical background of surface RREA observations and the reasons for their delayed recognition over many decades are detailed in Chilingarian et al. (2025). In this section, we describe TGE research details, comparing RREA energy spectra modeled with the CORSIKA code and TGE observations. We used the CORSIKA code (Heck et al., 1998), version 7400, which includes electric field effects in particle transport (Buitink et al., 2010).

The growth of RREA increases the cloud‘s electrical conductivity. Studies in New Mexico (Marshall et al., 1995; Stolzenburg et al., 2007) have shown that lightning flashes occur after the AEF exceeds the RREA initiation threshold. RREA simulation codes do not include a lightning initiation mechanism. Therefore, one can artificially increase the electric field strength beyond realistic levels to generate billions of avalanche particles; however, this approach lacks physical justification. As a result, we do not test electric fields stronger than 2.2 kV/cm at altitudes of 3-6 km.

The RREA simulation used vertical seed electrons and a uniform electric field that exceeded the Eth by 10-30% at an altitude of 5,400 meters. However, applying a uniform electric field produces different excess percentages over 1at various altitudes, depending on air density. The seed electron energy spectrum was based on the EXPACS WEB calculator (Sato, 2016), following a power law with an index of -1.25 for energies from 1 to 300 MeV. The ambient secondary cosmic ray population at 5-6 km altitude contributed approximately 42,000 seed electrons per square meter per second with energies above 7 MeV, as estimated with EXPACS. During large TGE events, the typical distance to the cloud base is estimated to be 25–100 m (see Fig. 17 in Chilingarian et al., 2020). To study electron flux decay, particle propagation continues through dense air for an additional 100 m before detection. The simulations included 1,000 to 10,000 events for electric field strengths ranging from 1.8 to 2.2 kV/cm, as shown in Fig. 2. Electron and gamma-ray propagation was tracked until their energies dropped to 0.05 MeV. The number of RREA particles was recorded every 200 m in the AEF and every 25 m after leaving the AEF.
In Figure 2, we show the development of the RRE avalanches at different atmospheric depths and for various physically justified strengths of the intracloud electric field. The curves are recalculated using a single seed electron for easier comparison with experimentally measured intensities. For the lower electric field strengths (1.8 and 1.9 kV/cm), the RREA process attenuates before reaching the observation level at 3200 m, at a depth of 2100 m (see the red and yellow curves in Fig. 2).

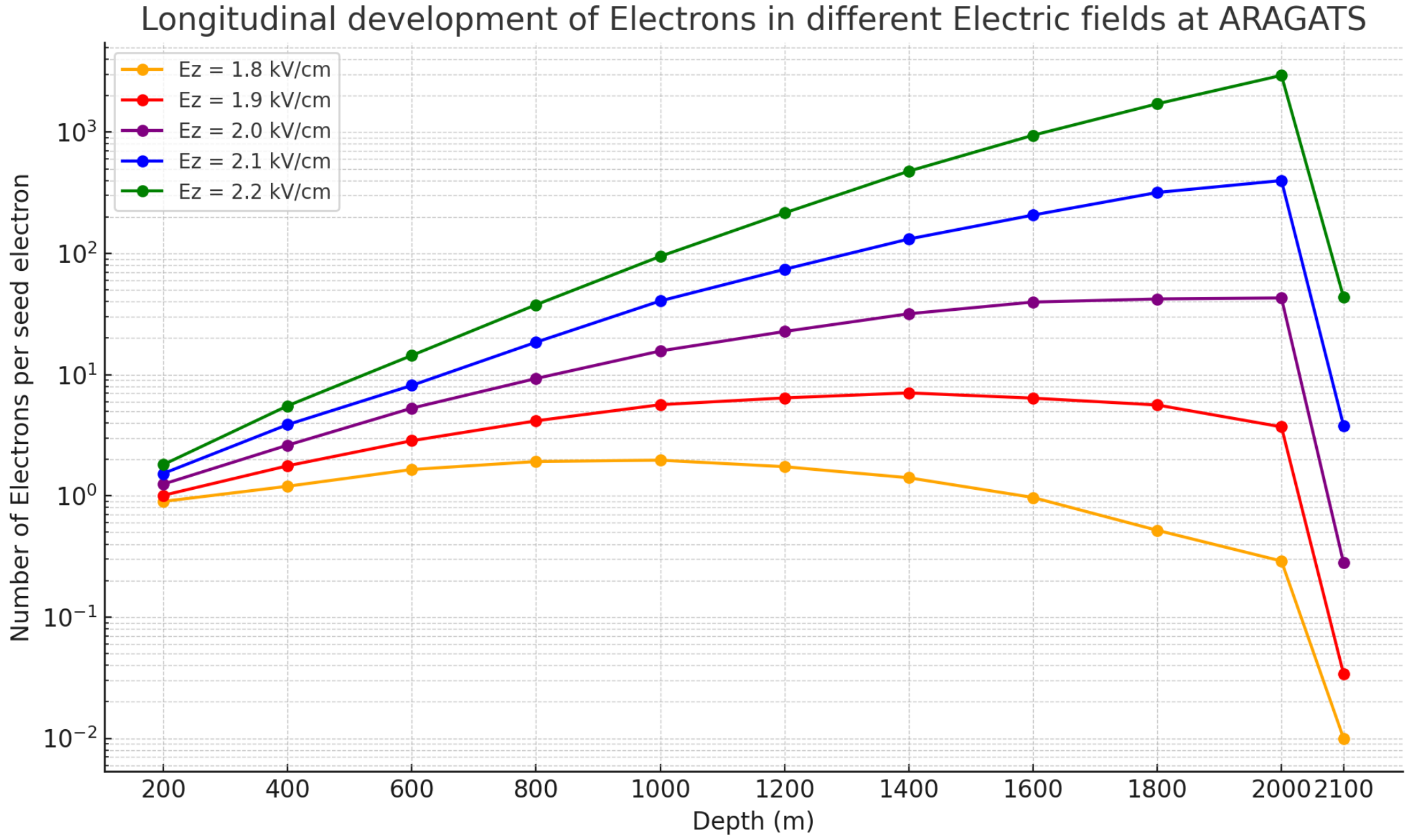

**Figure 2. Development of the RRE avalanche in the lower dipole. The avalanche began at 5400 m a.s.l. (0 meters depth), which is 2100 meters above the Aragats station. The number of avalanche particles is calculated every 200 meters. After exiting the electric field, the propagation of avalanche particles is tracked for an additional 100 meters before reaching the station, which is located at 2100 m depth.**

We continuously supported TGE measurements on Aragats with simulations of the RREA process in AEFs. Previously, we compared these simulations with the TGE recorded on June 14, 2020 (Chilingarian et al., 2021a). In this paper, we analyze the TGE that occurred on October 2, 2024, at 00:43. Particle detectors on Aragats registered a significant increase in particle flux count rate, as shown in Fig. 3. This was the first and only TGE of 2024 with a large electron content. The enhancement of low-energy gamma-ray and electron flux detected by the stacked 3-cm-thick, 1-m²-area STAND3 plastic scintillator (see Chilingarian and Hovsepyan, 2023, coincidence 1000, red curve) reached 225% (125σ). The "1100" coincidence of the same detector (green curve) selected electrons with energies above 20 MeV. The 20-cm-thick, 0.25-m²-area spectrometric SEVAN light scintillator (Chilingarian et al., 2024) detected electrons with energies above 10 MeV (blue). Gamma-ray peaks are typically much larger (black and red curves), reflecting the rapid attenuation of electrons upon leaving AEF.

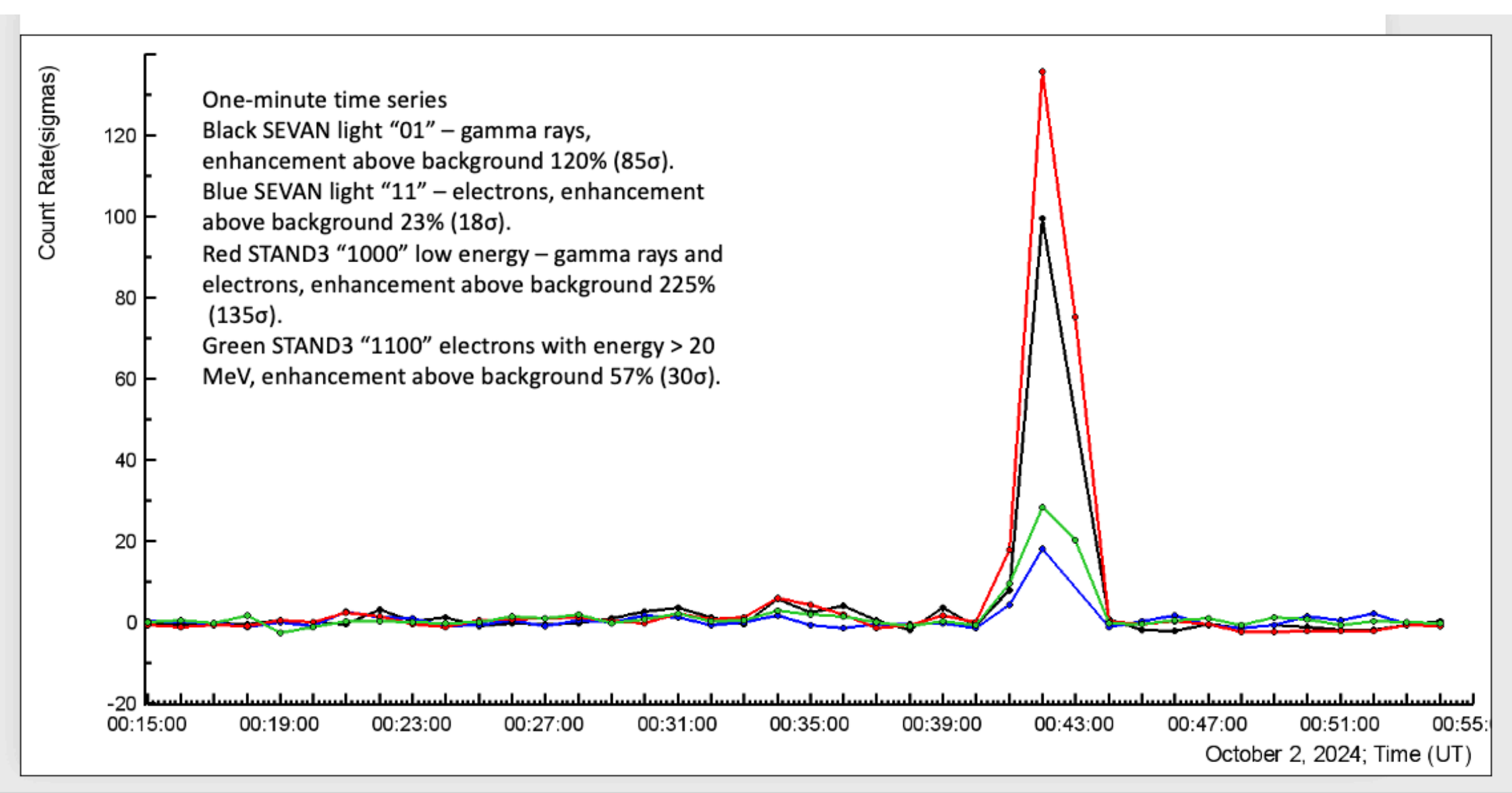

**Figure 3. One-minute time series of count rates for SEVAN light (black and blue) and STAND3 detectors (red and green).**

More details on TGE are shown in Fig. 4. A one-second time series of the count rate from a 1 cm thick, 1 m² outdoor scintillator on the roof of the GAMMA experiment's calorimeter shows nearly a tenfold increase (850%, 120$\sigma$). The black curve illustrates the NSEF, which is deeply negative during TGE. A broad peak started at 00:41:40, reached a maximum at 00:42:35, and then declined at 00:43:15 before immediately rising again until a cloud-to-ground (-CG) lightning flash abruptly stopped TGE at 00:43:36. The RREA electron flux in the cloud was intense enough to create an ionization channel in the lower atmosphere, providing a path for the lightning leader (see Chilingarian et al., 2017a).

In the inset of Fig. 4, we show the distance to the cloud base, estimated by the spread (difference between outside temperature and dew point), using the well-known meteorological approximation equation (Spread, 2025):

**H(m) ≈ (Air temperature at surface {◦C} − dew point temperature {◦C}) × 122 (1)**

The distance is minimal (60 m) at the flux maximum. This explains a 3.5-fold increase in gamma-ray flux compared to the electron flux, as measured by the SEVAN light spectrometer (5744 vs. 1660). Electrons were attenuated over 60 m before reaching the detector.

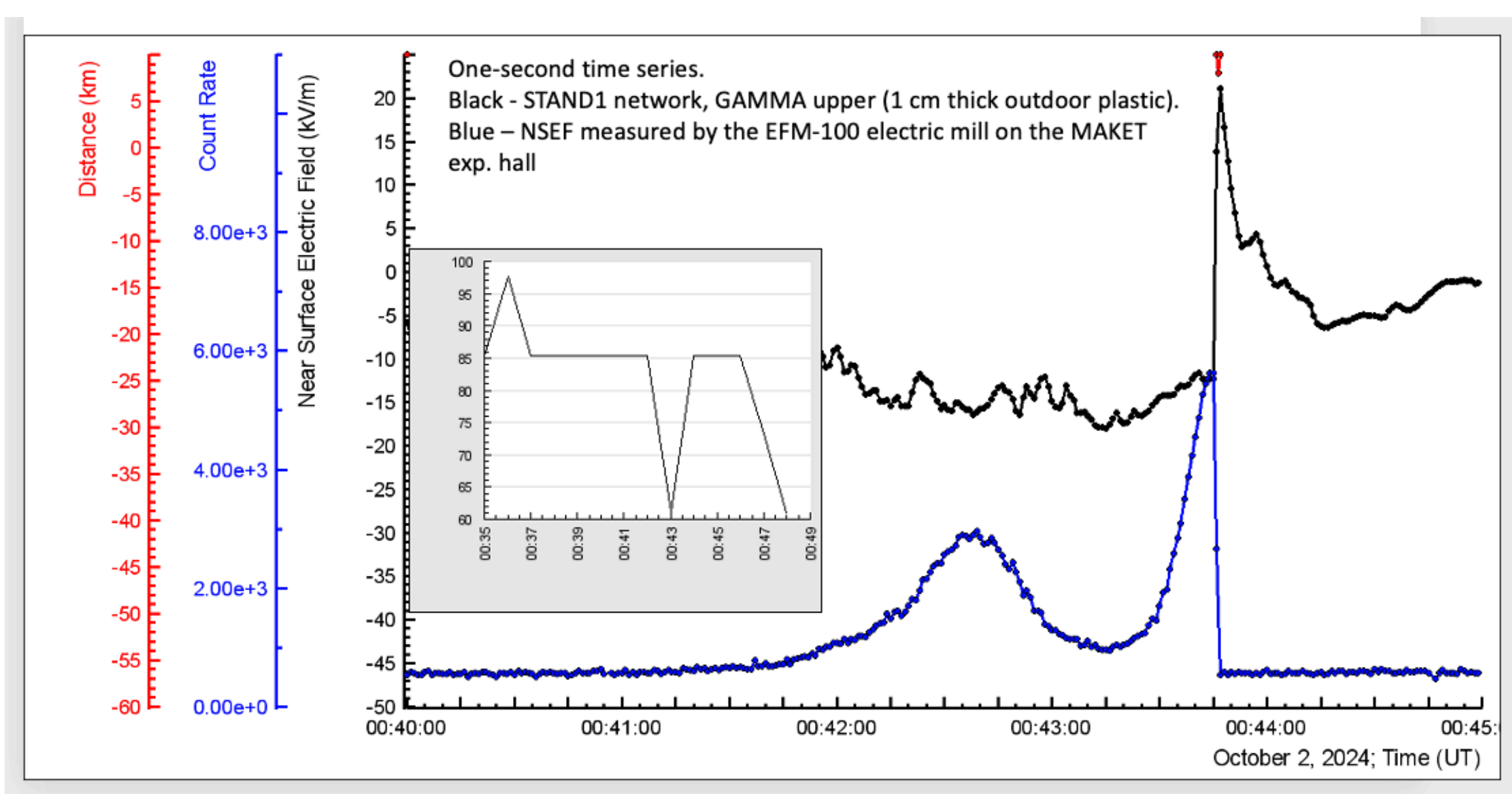

**Figure 4. One-second time series of the STAND1 detector (blue) and measurements of NSEF (black curve). In the inset – distance to the cloud base.**

In Figure 5, we present the differential energy spectrum of electrons measured above the roof of the SKL experimental hall, where the SEVAN light spectrometer is positioned. The detector response was unfolded using a GEANT4-based response matrix, following the inverse-problem method outlined in Chilingarian et al. (2022). The spectrum is well described by a simple exponential function in energy and aligns with the form given by Eq. 17 in Dwyer & Babich (2011). The maximum electron energy reaches approximately 34 MeV.

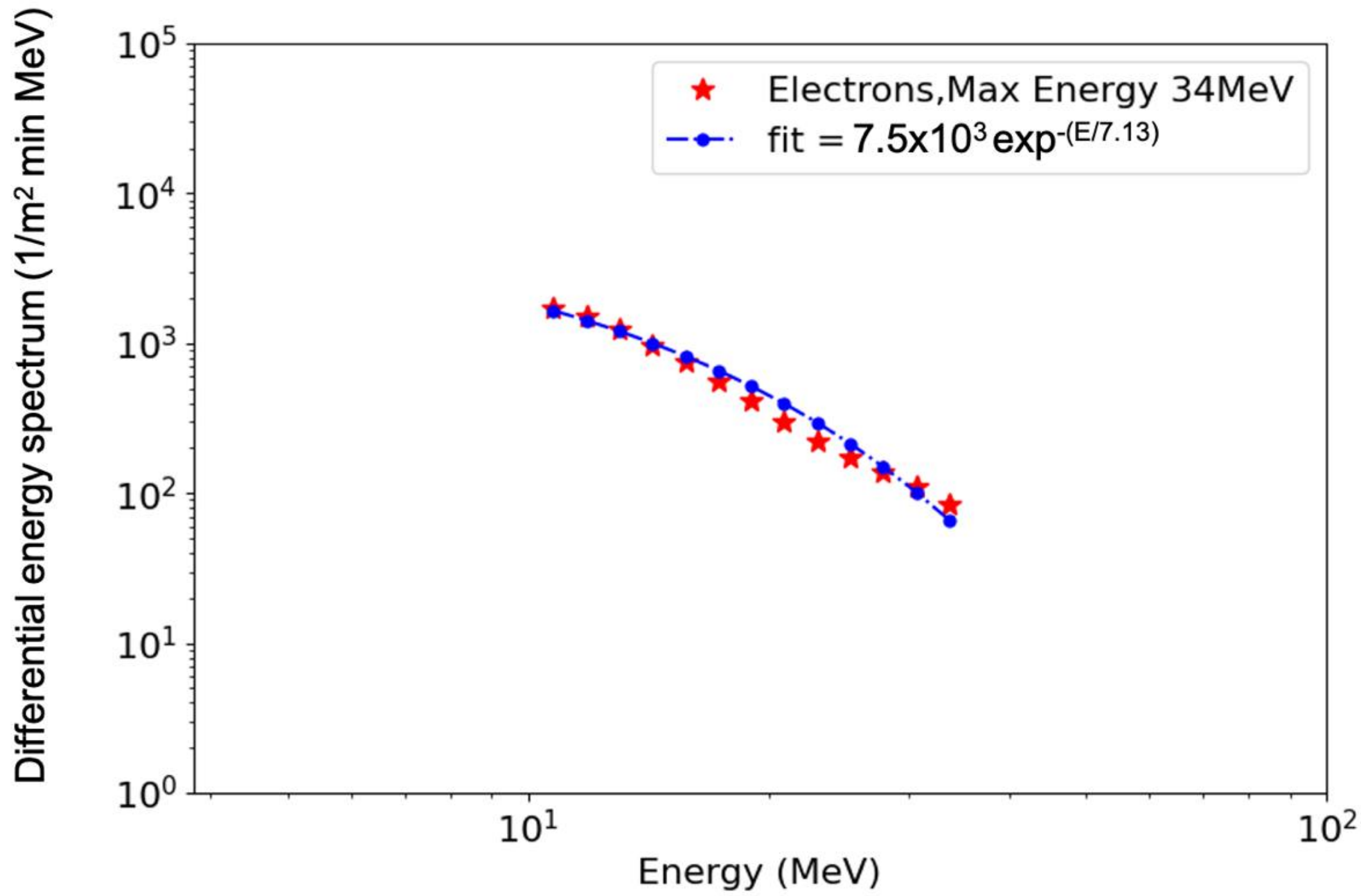

**Figure 5.** *Differential energy spectrum of TGE electrons above the roof of the SKL experimental hall (SEVAN light spectrometer).*

We integrated the recovered differential spectrum above 10 MeV to obtain the integral electron flux (counts $min^{-1}$ $m^{-2}$), which is directly comparable to CORSIKA simulations. The per-seed multiplication factors used in the comparison are taken from CORSIKA and represent the mean number of electrons with energies >10 MeV that reach the detector level for each injected seed electron at 5.4 km a.s.l. Simulations were run for several intracloud field strengths Ez and accelerating-field extents L, with avalanches tracked down to the Aragats station (3.2 km). The measured and simulated integral fluxes (count rates) are compared in Table 1. We include the TGE of 14 June 2020 (a moderate event) and 2 October 2024 (a strong event).

**Table 1. Comparison of simulated and measured electron fluxes for 2 TGE events occurred on 14 June 2020 and 2 October 2024**

| Case | Field Ez (kV/cm) | AEF extent L (km) | Termination Above Detector (km) | Measured I(>10 MeV) (counts $min^{-1}$ $m^{-2}$) | Simulated I(>10 MeV) (counts $min^{-1}$ $m^{-2}$) |
|---|---|---|---|---|---|
| 2020 small TGEs | 1.8–1.9 | ≈1.0–1.5 (≤1.2 for 1.8) | 0.10–0.20 | ~2–5×$10^3$ | ~2–4×$10^3$ |

| 2024 strong TGE (best) | ≈2.0 | ≈2.0 | 0.06 | $1.31\times10^4$ | $1.35\times10^4$ |
|---|---|---|---|---|---|
| 2024 strong TGE (alt.) | ≈2.1 | ≈1.2–1.4 | 0.06 | $1.31\times10^4$ | $1.28\times10^4$ |

The comparisons of simulated data shown in Fig.1 with the experimentally measured energy spectrum in 2020 (Fig. 4 of Chilingarian et al., 2021a) and 2024 (Fig. 5) yield the following conclusions:

- 2020 TGE. Measured and simulated integral fluxes agree within a factor of ~1.2–1.5, which is comfortably inside expected variability from atmospheric conditions and instrumental uncertainty. The accelerating field was (1.8–1.9 kV/cm) and elongation (1.0–1.5 km), with the lower boundary 100 m above the detectors. For such not-very-large TGEs, the AEF range is under ~1.5 km and the potential drop is under ~250 MV.
- 2024 strong TGE. Simulations with Ez = 2.0–2.1 kV/cm and L = 2.0 km reproduce the measured integral flux with high accuracy. The potential drop can exceed ~400 MV.
- Higher fields (e.g., 2.2 kV/cm) overshoot the intensity and are inconsistent with the data.

Therefore, both the 2020 and 2024 TGEs support RREA originating in the lower dipole, starting at 5.4 km a.s.l., and extending to the lower AEF boundary, located tens to hundreds of meters above the detectors. The shorter termination height inferred for 2024 naturally explains its much higher flux. The shorter termination heights in 2024 compared to 2020 account for the significantly larger fluxes observed in 2024. This agreement supports the interpretation that TGEs at Aragats are produced by RREAs initiated in thunderclouds above the station, with their strength and vertical extent governing the resulting ground-level flux.

Modelling AEF and the development of RREA within it were based on many simplifications. The strengths and spatial extent of the electric field, cloud height, and seed electron energy spectrum are assumed in the simulation. However, they are in overall agreement with in situ measurements, but can significantly deviate from the conditions of the particular thunderstorm that gave rise to the detected TGE event. Therefore, the energy spectra obtained in simulation trials provide only an overall estimate of the particle yield on the surface. We also made some assumptions about the termination of the AEF, relating it to the spread. Although this method is widely used in meteorology, we cannot one-to-one identify the AEF termination with cloud base height. Therefore, we develop an empirical method based on the maximum energies of the measured electron and gamma-ray spectra (see Chilingarian et al., 2022).

Using the Mendeley dataset (Chilingarian and Hovsepyan, 2021), we recovered the electron differential energy spectra by analyzing 16 TGE events and estimated the heights of the field termination above the ground. The distance at which the strong accelerating field is terminated (free passage distance, FPD) is determined using CORSIKA simulations by an empirical equation tuned on available electron spectra (Spread, 2025):

$$FPD\ (meters) = (C1*E^{\gamma}_{max} - E^{e}_{max})/C2 \quad (2),$$

We identify the highest energies of electrons and gamma rays from recovered energy spectra. Coefficients C1 and C2 are 1.2-1.4 and 0.2, respectively. TGE simulations indicate that the maximum energy of electrons leaving the electric field is 20-40% higher than that of gamma rays. Therefore, we estimate the maximum energy of electrons exiting the field as C1*E_max. We performed multiple simulations of electron-gamma ray avalanches to verify the accuracy of equation (2) and check for potential methodological errors. We record particle energies and solve the inverse problem to recover RREA characteristics from measured TGE data. We use CORSIKA simulations with different electric field strengths and termination heights to achieve this. Subsequently, we follow all experimental procedures on the obtained samples to estimate the maximum energies of electrons and gamma rays. Then, we calculate the FPL parameter using equation (2) and compare it to the "true" value from the simulation. Based on this comparison, we estimate the method's mean square deviation (MSD) as 50 meters, as shown in Table 12 of Chilingarian et al. (2022, supplementary materials). Using the maximum energies of gamma rays (28 MeV) and electrons (30 MeV), we estimate the FPD on October 2, 2024, to be 50 ± 50 meters. This estimate is consistent within one standard deviation with the value derived from "meteorological" data, which is 60 m, according to the spread (Eq. 1).

Another piece of evidence for the strong AEF on October 2, 2024, is the significant depletion of the muon flux, which supports the idea of an extended electric field above the detector. The same electric field that accelerates electrons also reduces positive muons, and because of a charge ratio of 1.2–1.3, the muon counts in the scintillator show substantial depletion; see Fig. 6. It also highlights that TGE can be used to estimate the maximum electric field in the thundercloud (see Fig. 4 and Table 2 in Chilingarian et al., 2022).

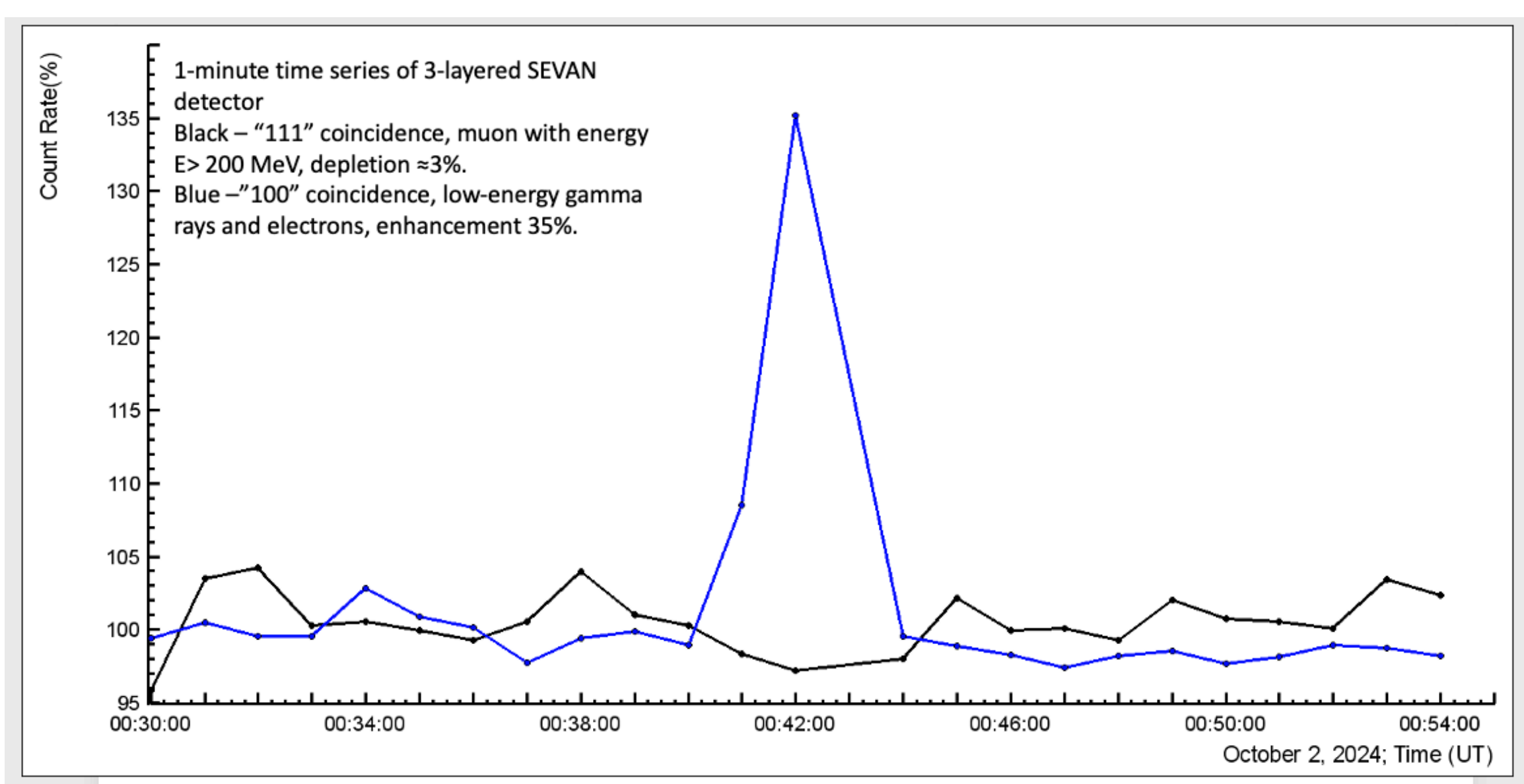

**Figure 6. One-minute time series of the SEVAN-Aragats detector, showing black low-energy gamma rays and electrons, and blue high-energy muons.**

Based on key measurements of the TGE electron and gamma-ray energy spectra, along with observations of neutron and muon flux modulation effects on Aragats and worldwide (Chilingarian et al., 2025), we conclude that the RREA model with EAS seeds fully explains electron-photon avalanches in the lower atmosphere. A comprehensive model of enhanced particle fluxes in the upper atmosphere and space has yet to be developed. The biggest challenge is identifying the source of seed electrons. EAS electrons and positrons move downward, while generating a population of upward-moving electrons, which presents a significant challenge.

## 3. Enhanced particle fluxes above thunderstorms: the seed electrons problem

Historically, the earliest evidence of high-energy radiation in the upper atmosphere comes from decades of balloon and aircraft observations of intense gamma-ray bursts at altitudes of 8-15 km (see discussion and references in Ostgaard et al., 2019). These measurements, often interpreted as manifestations of the MOS process (Chilingarian, Mailyan, and Vanyan, 2012), assume that the AEF did not exceed the runway threshold at altitudes of 15-20 km. However, the source of seed electrons remains unclear.

In the lower dipole, the ambient downward-moving cosmic-ray electrons align with the downward-directed electric field and easily initiate RREA. In the upper dipole, however, the field points upward, opposite to the momentum of the ambient cosmic-ray electrons; CORSIKA simulations show that such particles do not supply significant numbers of upward >1 MeV seeds. Therefore, the seed mechanism for upper-dipole RREA is fundamentally different from that at lower altitudes.

There is extensive evidence, ranging from early balloon and NASA's F-106 jet flights (Parks et al., 1981; McCarthy et al., 1985) to recent aircraft-based observations (Kelley et al., 2015; Kochkin et al., 2017; Ostgaard et al., 2019), that lightning abruptly halts particle fluxes in the upper atmosphere. This suggests a relationship between the two phenomena and warrants further investigation into RREA seeds in lightning initiation. It has been proposed that a small-scale electric field at the lightning leader tip can be strong enough to initiate RREA and produce abundant seed electrons, resulting in an extremely bright source with up to $10^{20}$ fluence (cold runaway model, Celestin & Pasko, 2011). In 2012, the relation between lightning and TGF was thought to be firmly established (Dwyer et al., 2012): "TGFs are produced in what are structurally normal IC flashes during the period when the initial negative polarity leader travels upward from the main negative charge layer to the upper positive layer. This occurs during the first 5–10 ms of the lightning flash, but distinctly after the flash initiation. TGFs occur during the ascent of this upward leader before it reaches and expands into the upper positive charge layer. Inferred altitudes of the leader tip at the time of TGF generation have ranged from approximately 11 to 15 km.

In 2024, this relationship seemed to be smeared (Dwyer, Rassoul, 2024): "Currently, TGFs are thought to be produced inside thunderclouds during the initial stage of upward positive intra-cloud (IC) lightning. However, the relationship between lightning and TGF production is unclear, and it is also not well understood why some lightning produces TGFs while others do not. In addition to spacecraft observations, X-rays and gamma rays are emitted by thunderclouds, as observed using both in situ and ground-based techniques. These emissions often form gamma-ray "glows" lasting seconds to minutes (thunderstorm ground enhancements when observed from the ground). Many of these observations exhibit energy spectra similar to TGFs, which extend into the multi-MeV range, indicating a similar source mechanism. In 2025, it was firmly declared that "thunderstorms may generate intense gamma-ray emissions without lightning" and TGFs can be created by "thunderstorm electrification alone, without the presence of lightning…" (Dwyer, 2025).

Thus, recent measurements and reanalysis of TGF catalogs do not support the "lightning" scenario of TGF origination (see review Chilingarian, 2024, and references therein). TGFs, measured alongside lightning flash detections by ASIM instruments, prove that there are no observed events where the optical pulse onset precedes the onset of the TGF. The median determined delay between onsets is 190 µs (Skeie et al., 2022). In the multi-pulse patterns of TGFs registered by ASIM (Fig. 6.6 of Fuglestad, 2023), we can see that the first TGF detected at 18:02:25 on July 5, 2021, smoothly finished, and no optical images of the atmospheric discharges were observed. The second TGF, which occurred 2 ms later, was terminated by a lightning flash, as seen in both the abrupt termination of particle flux and the optical signal. Zhang et al. (2021) demonstrated that gamma rays are produced several milliseconds before a narrow bipolar event, which often marks lightning initiation. Analysis of four TGF catalogs from different instruments revealed that a significant proportion of TGFs lead to increased lightning activity detected in radio waves (spherics) between 150-750 ms after TGFs occur (Lindanger et al., 2022). Furthermore, in a recent paper (Gourbin and Celestin, 2024), the maximum achievable number of electrons from “cold runaway" was limited to $10^{17}$.

The Airborne Lightning Observatory (ALOFT, Ostgaard et al., 2024) conducted cutting-edge experiments in 2023, measuring multiple gamma glows from tropical thunderstorms. ALOFT payload consists of five spectrometers, 30 photometers, three electric-field sensors, two radars, and two passive radiometers. Particle fluxes were monitored during flight, and gamma-ray-glowing clouds were identified in real-time to facilitate return and continuation of measurements. ALOFT detected more than five hundred gamma-ray glows during nine of the ten flights, showing that thunderclouds can emit gamma rays for hours and over huge regions (see Fig. 1 of Marisaldi et al., 2024). Glows were detected repeatedly following consecutive passages (aircraft revisited flaring thunderclouds after notification of mission physicists) over the same thundercloud system, covering $\approx 10^4$ km².

Correlated measurements of ALOFT and ASIM reveal several gamma-ray fluxes observed by ALOFT but not by ASIM during the ISS overpass. The authors (Bjørge-Engeland et al., 2024) conclude that the overwhelming gamma glow population directed to space from tropical thunderclouds is too weak to be observed from space. The source photon brightness of gamma glows is several orders of magnitude lower than what is usually attributed to TGF observed by orbiting gamma observatories. Thus, TGFs comprise only a small percentage of gamma glows copiously observed above tropical storms, as numerous TGEs are observed on the Earth’s surface. In our opinion, gamma glows and TGFs may represent different observational manifestations of the same underlying high-field radiation process; only a very small fraction of the upward-directed “glow gamma rays” are favorably oriented to be detected by satellites at altitudes of hundreds of kilometers.

Thus, the ALOFT experiment provided groundbreaking insights, demonstrating that gamma glows and TGFs were immediately followed by Narrow Bipolar Events (NBEs), initiating significant lightning activity. These findings challenge prior hypotheses that lightning leaders supply the seed electrons for TGFs. Recent measurements from ALOFT and long-term nanosecond-scale monitoring of atmospheric discharges and particle fluxes at Aragats strongly support an alternative hypothesis: relativistic runaway electron avalanches precede and initiate the lightning leader (Chilingarian et al., 2015, 2017b). The combination of gamma-ray imaging and lightning interferometry proposed by the ALOFT continuation can directly test and validate this transformative hypothesis.

The ALOFT findings, which demonstrate the role of thunderclouds as huge natural particle accelerators, confirm ground-based observations obtained by continuously monitoring particle fluxes and electric fields within large areas around Aragats Mountain (Chilingarian et al., 2022; Chilingarian et al., 2024c). Thus, the joint study of TGEs and gamma glows will enhance fundamental knowledge of high-energy physics in the atmosphere (HEPA), clarifying key issues related to particle acceleration and lightning initiation while bridging the gap between processes in the lower and upper atmosphere (Dwyer et al., 2012).

Observations of ALOFT (Marisaldi et al., 2024; Ostgaard et al., 2024), conducted 1-2 kilometers above gamma-ray sources, yield groundbreaking findings that demonstrate the existence of extensive and prolonged gamma-ray emission regions, contradicting earlier theoretical models used for 30 years to explain TGFs. These newly characterized phenomena, such as Flickering Gamma-ray Flashes (FGFs) and Glow Bursts (GBs), challenge models featuring intense, distributed gamma-ray sources and open a fresh frontier in the RREA/gamma glow models, making them similar to the RREA/TGE model.

In the next section, we will show how the hybrid Compton-redirected gamma rays' mechanism (the dual-stage model, DSM)—where strong RREAs in the lower dipole produce upward Compton-redirected gamma rays—naturally accounts for both the flux level and the vertical size of gamma glows observed by ALOFT.

**4. Dual-stage model: RREA mechanism for upward gamma-ray glows**

While it is generally understood that the physical mechanism responsible for atmospheric radiation is RREA, the source of seed electrons varies. Historically, TGFs are modeled with highly intense local gamma sources, with fluence reaching up to $10^{19}$. In contrast, TGEs are modeled with seeds from EAS electrons, providing a well-measured, stable, altitude-dependent flux over large volumes of thundercloud systems. Persistent, upward-directed gamma-ray glows observed by ALOFT and balloon experiments—covering areas greater than or equal to 10,000 $km^2$ and lasting from tens of minutes to hours—cannot be explained by short, localized very bright discharge scenarios (e.g., classic TGF models). Therefore, we propose a dual-stage, vertically coupled mechanism in which a lower-dipole RREA supplies photons that seed an upper-cloud RREA, thus naturally producing a sustained gamma flux that extends into space across large areas (~$10^4$ $km^2$), spans multiple kilometers vertically, and persists for several tens of minutes.

The issue of seed formation for the upward-moving avalanche has been essential since the discovery of the TGF in 1994 (Fishman et al., 1994), and it has become even more pressing following the detection of intense gamma-ray flybys with the ALOFT experiment (Marisaldi et al., 2024). However, unlike the lower dipole, the physics of electron acceleration in the upper dipole remains not fully understood. In this paper, we will analyze in detail the physical processes that seed relativistic runaway electron avalanches (RREAs) in the upper dipole, leading to particle bursts in the upper atmosphere and space. To prevent misunderstandings about the codes used and the interpretation of simulation results, we will add simplified theoretical considerations of particle interactions in the upper atmosphere to the simulation results.

In the CORSIKA code, the atmospheric electric field (AEF, Ez) negative sign indicates electron acceleration toward open space, while the positive sign indicates electrons accelerating toward Earth's surface (physical notation). The notation for atmospheric electricity used in geophysics uses reverse notation: positive AEF indicates accelerated electrons toward open space, and negative AEF indicates electrons moving toward Earth's surface (atmospheric electric field notation). In this section, we will use the physical notation accepted by CORSIKA authors.

The downward ambient cosmic ray (CR) flux provides electron seeds for the RREA developing in the lower dipole. For RREA to develop in the upper dipole, seeds must move upward, opposite to the CR flux. CR electrons in the upper dipole will be stopped by AEF; however, positrons will be accelerated and multiplied if AEF exceeds the runaway threshold. Additionally, gamma rays can travel freely through the AEF of either polarity. We will consider all channels that supply upward-moving electrons for RREA development in the upper dipole. MeV positrons can trigger RREA avalanches, and gamma rays from these avalanches can produce electron-positron pairs ($\gamma \to e^- + e^+$), resulting in some upward electrons. CR gamma rays can also generate electrons through pair production. Furthermore, bremsstrahlung and Compton-scattered electrons can be directed upward. We will examine all sources of upward electrons that serve as seed electrons entering the bottom of the upper dipole.

For calculation and simulation purposes, we should choose realistic parameters based on the RREA process in the atmosphere, justified by both theoretical considerations and extensive experience with RREA detection in the lower dipole.

## 5. RREA development calculation and simulation

We adopted the tripole structure of the AEF (Kuettner, 1950). The tripole was positioned within the 5.3-7.3 km range (lower dipole) and the 7.3-9.3 km range (upper dipole). A uniform electric field was applied within these dipoles. Because the AEF is quite long, we divided the 2 km into 10 slabs of 200 m each for illustrative purposes and calculated the seed yield for each. The consideration of RREA multiplication is based on the concepts of e-folding length and multiplication rate. The e-folding length λ is the distance over which the number of runaway electrons increases by a factor of e. Therefore, the RREA multiplication factor M can be approximated by:

$$M = \frac{N(z)}{N_0} = e^{z/\lambda}.$$

M quantifies the avalanche growth rate and depends heavily on: Ez, the AEF strength, and $E_{th}$, the runaway threshold field, which depends on air density ρ. Dwyer (2003, 2012) and Babich (2007) derived a convenient parameterization.

$$\lambda = \frac{73}{\left(E_z - E_{\mathrm{th}}(\rho)\right)},$$

where: $E_{th}(\rho)$, or $E_{th}(h)$ is the threshold (critical) field for runaway in kV/cm, ρ is air density, corresponding to height h, and λ is the e-folding length in meters.

The total multiplication along AEF length L is approximately:

$$M = \exp\,(L/\lambda).$$

The dependence of the threshold $E_{th}(h)$ on altitude was determined using the CORSIKA code, performed for several heights across multiple simulations with stepwise AEF growth; the threshold was set to the AEF at which particle exponential growth was detected (Chilingarian et al., 2025). In Fig. 7, we show this dependence from 2 to 10 km (black curve). The red curve shows a 30% increased AEF relative to Eth, at which RREA will be mature, and lightning flashes still will not quench AEF. We perform calculations for various values of Ez > Eth, aiming to stay within physically justifiable limits. Corresponding e-folding length λ is shown by the dashed blue line. Recall that after a 30-40% increase above the threshold, lightning flashes typically quench the potential difference and terminate RREA (Stolzenburg et al., 2007). Previously, we estimated the maximum electric field (potential drop) on the mountain tops of Lomnicky Stit (Chilingarian et al., 2021b) and Aragats (Chilingarian et al., 2021c), with results in good agreement with direct measurements. The inset shows the numerical values, including the multiplication rate M, at heights of 7.3 and 9.3 km.

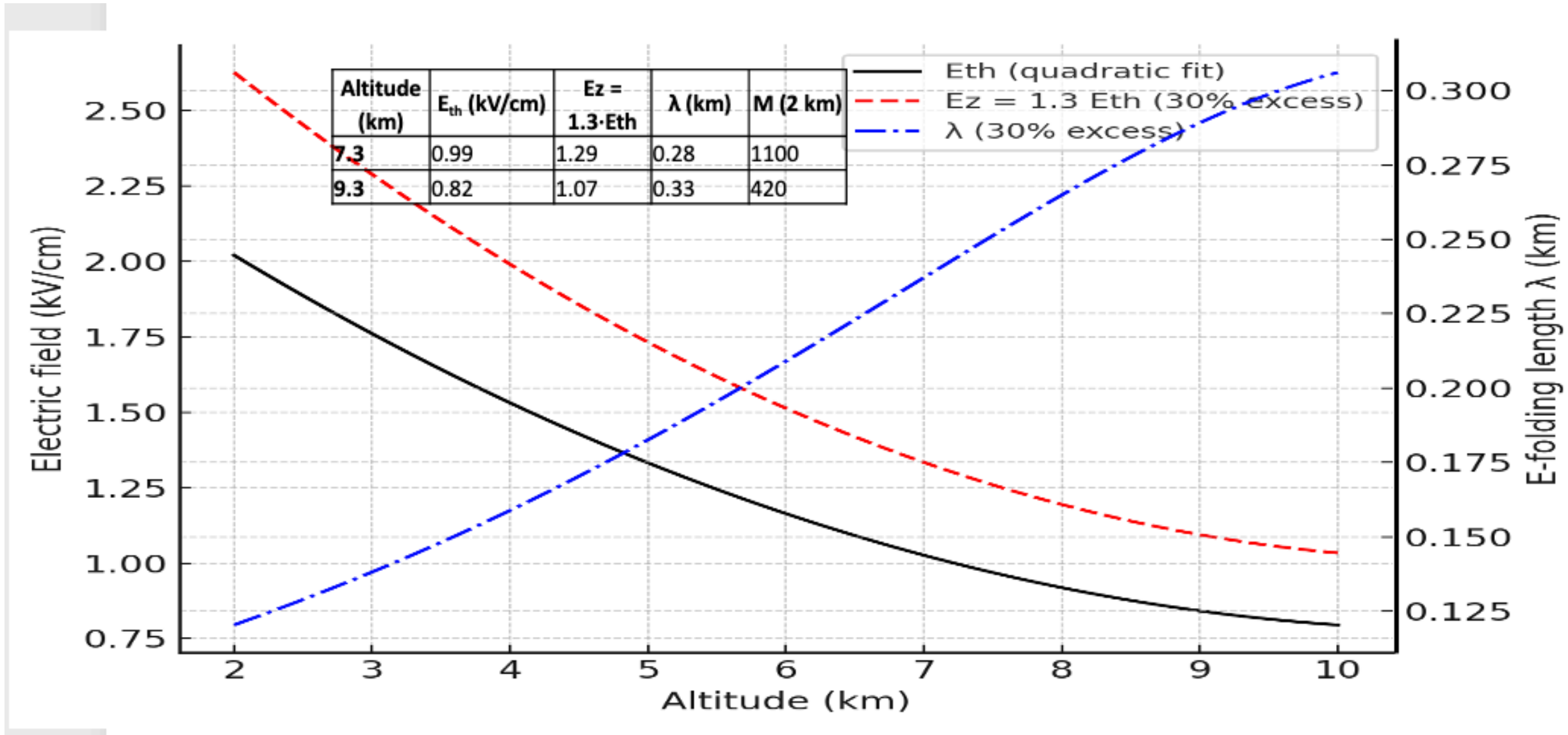

| Altitude (km) | $E_{th}$ (kV/cm) | Ez = 1.3·Eth | λ (km) | M (2 km) |
|---|---|---|---|---|
| **7.3** | 0.99 | 1.29 | 0.28 | 1100 |
| **9.3** | 0.82 | 1.07 | 0.33 | 420 |

**Figure 7. Altitude dependence of the threshold AEF and e-folding length. In the inset, we show a 30% enlarged $E_z$ used in calculations and the avalanche growth factor M in 2 km of field.**

In Figure 8, we illustrate the growth of RREA particles as AEF increases. Clearly, this growth is not physically justified, and, moreover, as we will demonstrate later, it does not lead to an increase in gamma rays, as indicated by the red line.

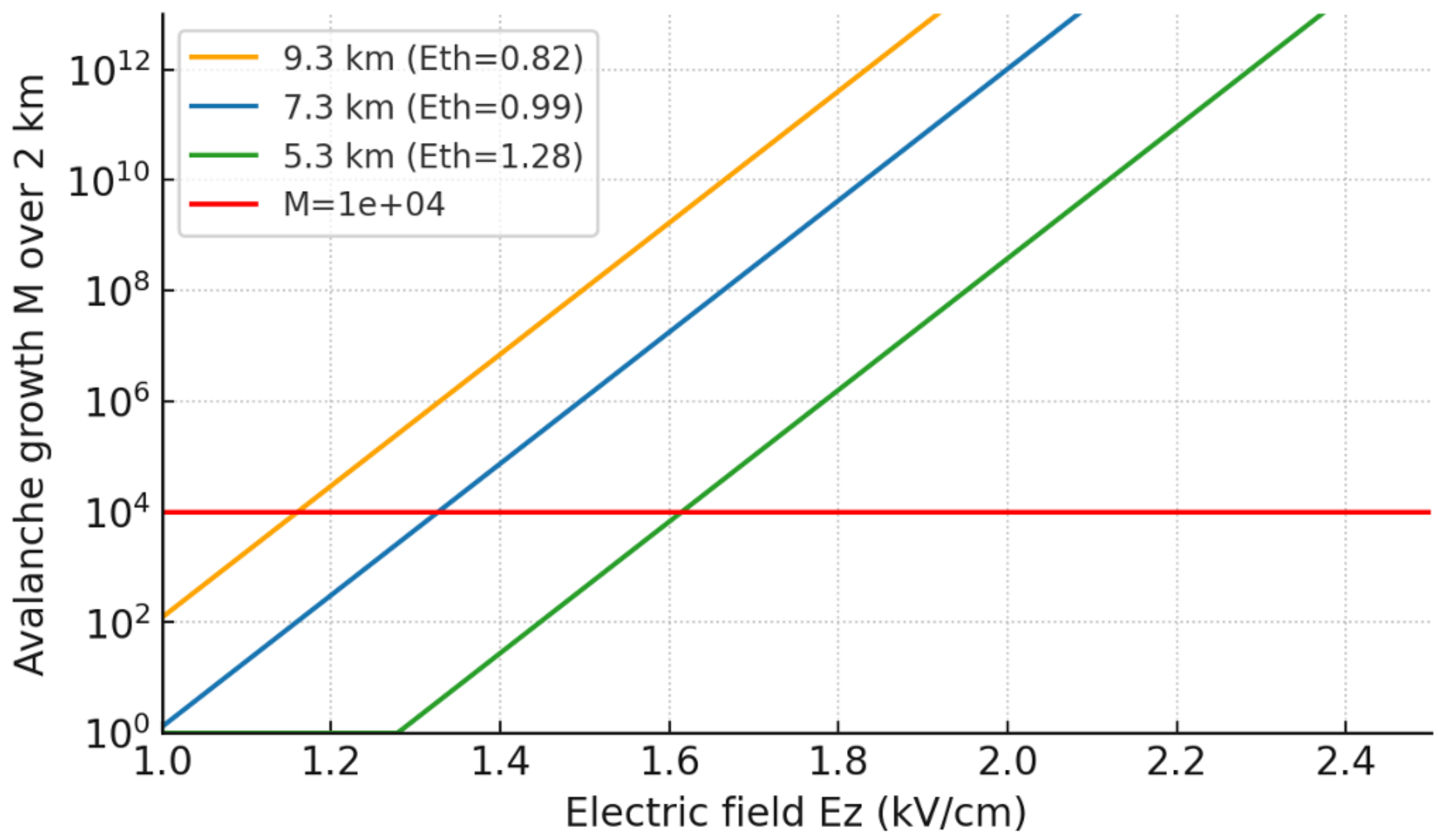

**Figure 8. RREA multiplication factor (avalanche growth) depends on Ez for different altitudes. The red line shows a limit on multiplication rates set to $10^4$.**

The limit for avalanche multiplication is $10^4$ based on our experience with TGE detection in the lower dipole (Chum et al., 2020; Chilingarian et al., 2024a). However, to simulate upgoing seeds, we used much higher AEFs here, as other researchers have, to demonstrate that even with very large, physically unjustified strengths, the number of upgoing seeds remains negligible. We simulate interactions of gamma-ray and positron beams with air and analyze the resulting upward lepton and gamma-ray fluxes. A total of 1,000,000 positrons and gamma rays are injected vertically downward from altitudes of 9.3 km and 7.3 km into a 200 m air slab ending at 9.1 km and 7.1 km. Their initial energy distributions follow power laws $dN/dE \propto E^{-1.125}$ and $E^{-2}$, spanning energies between 1 and 100 MeV. All gamma rays emitted and scattered upward, reaching 9.3 km and 7.3 km, are recorded and analyzed. Several earlier studies investigating the possibility of upward-directed runaway electrons used much stronger atmospheric electric fields. Dwyer (2003, 2007, 2012) employed fields of 2.5–3.0 kV/cm to explore relativistic feedback and the limiting behavior of upward-propagating runaway electrons above thunderstorms. Similar field strengths were used by Babich (2007; Babich et al., 2010), who applied 3–3.5 kV/cm to study extreme avalanche and positron feedback scenarios. These values significantly exceed realistic intracloud AEFs at 7–12 km altitude (see Figs 7 and 8), but were intentionally adopted to test whether even such exaggerated conditions can produce appreciable upward-moving seed fluxes. Consistent with these works, our simulations include cases with 2.1 and 2.4 kV/cm downward-directed fields to demonstrate that, even under extreme assumptions, the number of upgoing gamma rays and leptons remains negligible.

To track both upward and downward particle fluxes, we use CORSIKA 8 (Engel et al., 2019), built on the well-established CORSIKA 7 foundation. It extends capabilities beyond air showers to include particle cascades in various media, such as by introducing AEFs, and allows us to observe RREA cascades developing simultaneously in opposite directions within a single dipole. In the upper dipole, ambient CR flux positrons are accelerated downward in the AEF extended from 9.3 to 7.3 km, while backscattered electrons are accelerated upward, reaching 9.3 km. In the lower dipole, electrons from the ambient CR flux are accelerated downward in the AEF from 7.3 to 5.3 km, and positrons generated in air through pair production are accelerated upward to 7.7 km. In both directions and for both dipoles, positrons and electrons can develop RREA, producing numerous gamma rays that travel in both directions. Those with energies above 1.022 MeV can create electron-positron pairs, enabling backscattering of leptons.

In Figure 9, we display the energy release histograms of backscattered gamma rays from the downward positron and gamma ray CR fluxes for two levels of extreme electric fields, 2.1 and 2.3 kV/cm. Red lines show the fits for low- and high-energy ranges separately. The annihilation line at 511 keV is present in all four frames, with varying intensities. Other features in the spectra include bumps around 200 keV in frame c and near 1 MeV in frame d.

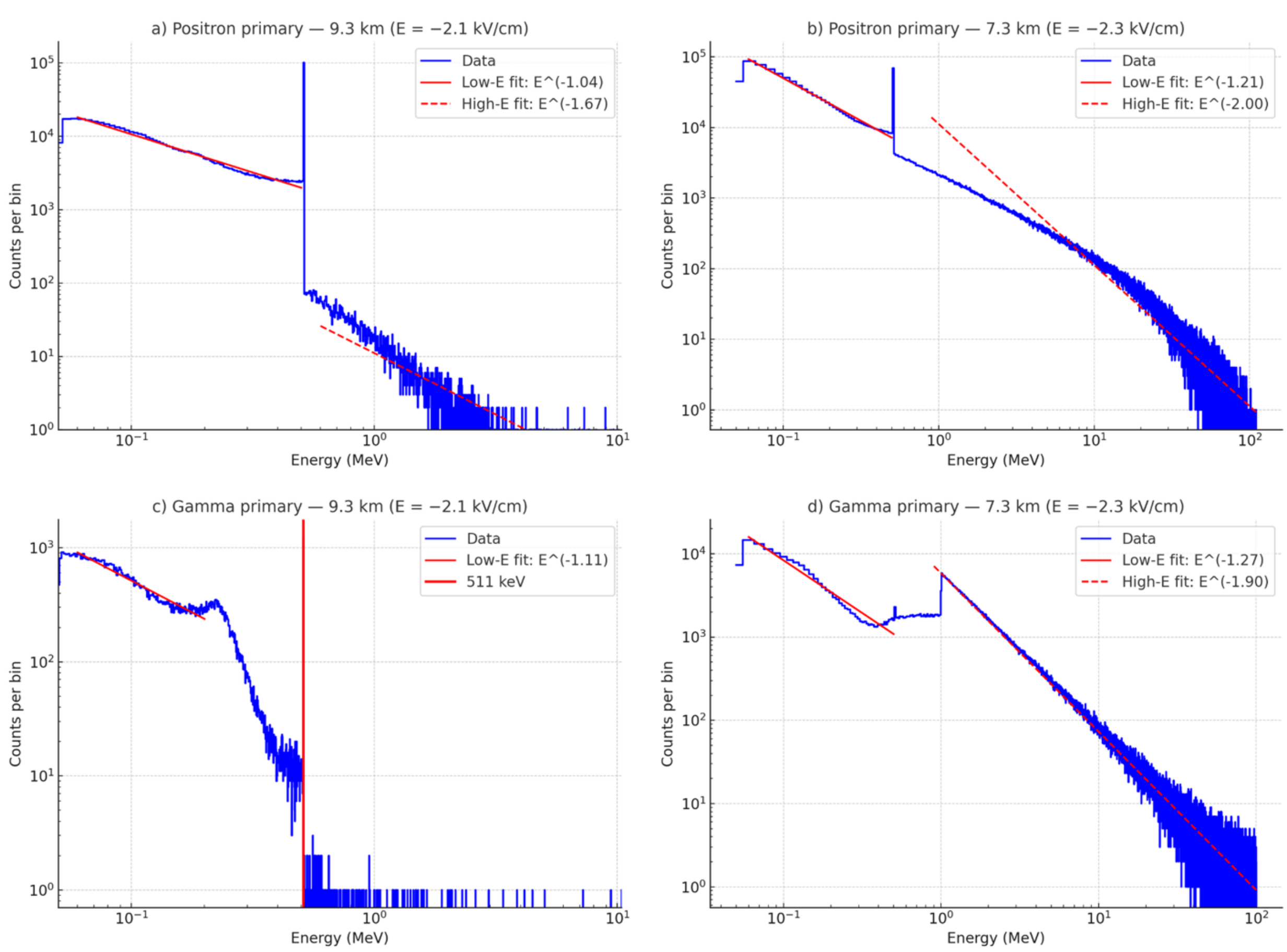

**Figure 9. Energy release histograms of backscattered gamma ray yield from incident gamma-ray and positron ambient fluxes at 7.3 and 9.3 km traversing 200 m of AEF.**

At energies below 0.5 MeV, the observed gamma-ray spectra are shaped by the combined effects of Compton scattering and bremsstrahlung. In this sub-MeV range, photons from bremsstrahlung and positron annihilation undergo multiple Compton scatterings in air. Each scattering lowers the photon energy and changes its direction, producing a nearly isotropic diffuse flux directed upward. This process, known as Comptonization, converts an initially hard bremsstrahlung spectrum into a quasi-flat continuum with an effective slope of about –0.9, matching the expected behavior of partially degraded bremsstrahlung in a low-density target (Pozdnyakov, Sobol & Sunyaev, 1983). The bump around 200–230 keV in frame c is the Compton backscatter peak produced when 511 keV annihilation photons scatter through ∼180° in air.

At 7.3 km, the higher density shortens the mean free paths of electrons and photons, increasing the likelihood that bremsstrahlung photons degrade to the sub-MeV range before escaping. This explains the observed increase in sub-MeV flux at lower altitudes. Pair production also influences this energy range through secondary cascades. High-energy photons (above 1.02 MeV) produce electron-positron pairs, which then emit secondary bremsstrahlung photons also in the MeV range. Many of these photons are later Compton-scattered down to hundreds of keV, adding to the sub-MeV continuum. Thus, the low-energy region results from the combined effects of primary bremsstrahlung degradation, reprocessing of annihilation photons, and downscattering of secondary photons from pair production.

A weak 511 keV feature and its Compton shoulder (notably in frame b) result from annihilation at rest, where roughly half of the photons escape upward. In-flight annihilation adds to the mid-range continuum (0.4–1 MeV) and merges with soft bremsstrahlung.

Downward-moving positrons emit bremsstrahlung photons that can reach tens of MeV and, in rare cases, escape into the upper hemisphere, creating the upgoing MeV continuum. A small number of electrons generated in pair-production events and occasionally directed upward also emit a broad bremsstrahlung spectrum extending to tens of MeV; since our goal is to account for all possible upward seeds, this weak channel is considered as well. The simulated MeV tails in the spectra (frames a, b, d) demonstrate this production of high-energy photons propagating upward. The sub-MeV continuum is produced primarily by Compton-degraded photons. In contrast, the MeV tail is replenished mainly by bremsstrahlung from positrons undergoing RREA in the strong upper-dipole field, with only a minor contribution from electrons created in rare pair-production events and accelerated in the opposite direction. The transition between the Compton-dominated sub-MeV continuum and the RREA-driven MeV tail appears as a mild spectral shoulder near ~1 MeV (frame d), most clearly visible in the positron-beam case, where the supply of radiating positrons is largest. Analysis of the raw spectra confirms that upward photon fluxes are higher at 7.3 km for positron primaries (compare frames c and d) in the sub-MeV region, and that both primary types exhibit stronger MeV-range emission at 7.3 km because the higher density enhances bremsstrahlung efficiency. The ~1 MeV shoulder therefore reflects the balance between Compton degradation and RREA-driven bremsstrahlung, with only a small contribution from pair-produced secondaries.

At 9.3 km, the runaway threshold is about 1 kV/cm, and the simulated field of 2.1 kV/cm is roughly double this value. At 7.3 km, the field (2.3 kV/cm) is approximately 75% above the local threshold, so both layers meet the RREA condition ($E > E_{th}$). Therefore, upward electron avalanching increases the high-energy tail of the spectra in both cases. However, despite the smaller $E/E_{th}$ ratio at 7.3 km, the resulting high-energy gamma-ray flux is much larger (see frames b and d). This happens because the air density and grammage within 200 m below 7.3 km are more than twice those at 9.3 km, which boosts the bremsstrahlung yield and pair-production probability. As a result, the denser layer produces a significantly higher MeV photon output even under a slightly less supercritical field.

Positron-initiated showers experience substantial bremsstrahlung losses while the positrons are accelerated in the downward electric field, which enhances secondary-electron production and increases the upward gamma-ray yield. Consequently, positron primaries generate more MeV photons and produce a harder high-energy spectrum than gamma primaries. The higher air density at 7.3 km further increases the probability of bremsstrahlung and pair-production interactions, so even a slightly less supercritical field produces a much stronger MeV photon output than at 9.3 km. Various scenarios leading to upward lepton fluxes are summarized in Table 2, where we report upward-moving lepton intensities above 50 keV and above 1 MeV for different AEF strengths, dipole geometries, and ambient cosmic-ray injection rates.

**Table 2. Summary of upgoing lepton yields for the single downgoing ambient positron and gamma ray.**

| **CORSIKA simulated upward particle yields (calculated from 1 million trials)** | **Ne+ (>1MeV)** | **Ne+ (>50KeV)** | **Ngamma (>1MeV)** | **Ngamma (>50KeV)** | **Ne- (>1MeV)** | **Ne- (>50KeV)** |
|---|---|---|---|---|---|---|
| **Upper_dipole_9.3 - 9.1 km observation level 9.35 km. -1.5kV/cm incident particle gamma.** | **0.000001** | **0.000001** | **0.00004** | **0.094** | **0** | **0.000011** |
| **Upper_dipole_9.3 - 9.1 km. observation level 9.35 km. -2.1kV/cm incident particle gamma.** | **0** | **0** | **0.00005** | **0.094** | **0** | **0.00005** |
| **Upper_dipole_9.3 - 9.1 km. observation level 9.35 km. -1.5KV/cm incident particle positron.** | **0.000004** | **0.000004** | **0.00020** | **0.0691** | **0.000069** | **0.00016** |

| **Upper_dipole_9.3 - 9.1 km. observation level 9.35 km. -2.1kV/cm incident particle positron.** | **0.000005** | **0.000005** | **0.00021** | **0.0697** | **0.000095** | **0.00018** |
|---|---|---|---|---|---|---|
| **Upper_dipole_9.3 - 7.3 km. observation level 9.35. -2.1kV/cm incident particle positron.** | **0.000007** | **0.000007** | **0.00121** | **0.375** | **0.000184** | **0.00082** |
| **Lower_dipole_7.3 - 5.3 km. observation level 7.35. +1.9kV/cm incident particle electron.** | **1.06** | **1.1** | **0.45** | **1.65** | **0.000222** | **0.000247** |

The data in the first five rows quantify the likelihood that a downward ambient cosmic-ray positron or gamma ray is converted into an upward-moving lepton or photon that returns to the top of the upper dipole and could, in principle, sustain the upward gamma-ray flux observed by particle detectors on balloons and aircraft. For each configuration, we inject 1 million primary particles and compute the mean number of upward particles per incident primary in a 200 m slab (9.3–9.1 km) and, for the extended case, along the full 2 km length of the upper dipole (9.3–7.3 km). We also compare a more realistic AEF of 1.5 kV/cm with an enhanced field of 2.1 kV/cm to evaluate how the electric field strength affects the probability of producing upward seeds.

The last row shows the **yield** of upward gamma rays and electrons reaching an altitude of 7.3 km from RREA initiated by ambient downward electrons in the lower dipole (5.3–7.3 km), with Ez = +1.9 kV/cm. Values greater than one in this row indicate that, on average, more than one upward lepton or photon is produced per incident electron, demonstrating efficient multiplication in the lower dipole. This contrast allows us to compare the efficiency of different mechanisms for supplying seeds to the upper dipole. In the next section, we use these results to evaluate whether an infinite feedback model (Dwyer, 2003; 2025; Babich, 2007) can plausibly explain the intense upward gamma-ray flux.

As shown in Table 2, a downgoing ambient cosmic-ray particle produces only $10^{-6}$ – $10^{-5}$ upward-moving MeV electrons capable of initiating RREA in the upper dipole. These values are far too small to support sustained RREA directed upward. Even when the electric field is raised to twice the local threshold (2.1 kV/cm), upward-moving MeV electrons are still essentially absent. A small number of sub-MeV electrons (<100 keV) appear due to Compton scattering of gamma rays, but their energies are well below the runaway threshold, and they cannot trigger an avalanche. Ambient positron flux produces more upward-moving photons than gamma primaries. This occurs because energetic positrons, when accelerated downward in the electric field, radiate bremsstrahlung over a broad angular distribution, and a non-negligible fraction of these photons naturally propagate into the upper hemisphere. Additional upward photons arise when these bremsstrahlung photons undergo Compton scattering in air. A similar mechanism operates in the lower dipole: downward-accelerated electrons emit bremsstrahlung, some of which is directed upward and can seed the upper dipole. In contrast, gamma primaries produce fewer upward photons because Compton scattering and pair production distribute energy more isotropically, yielding only a small upward component. Therefore, among ambient cosmic-ray particles, positrons are the most effective agents for producing upward-directed gamma rays, and we focus on positron-initiated RREA in the upper dipole.

For a 200-m layer, the upward gamma-ray yield at 1.5 and 2.1 kV/cm is almost the same because, over such a short distance, positron-initiated RREA has not yet formed. However, over the full 2-km length of the upper dipole, the avalanche becomes fully established, and the upward gamma-ray flux increases by about five times. In the lower dipole, the production of upward gamma rays is even higher—roughly four times greater—because the downward-moving electron beam interacts with air that is both denser and thicker. Although the AEF at 9.3 km significantly exceeds the runaway threshold, the air density is about 40% lower than at 7.3 km. This leads to two effects: first, the reduced collision rate results in fewer bremsstrahlung and pair-production interactions per unit path, lowering the number of upward photons. Second, the longer mean free paths at high altitude allow **photons** to escape the 200-m region without undergoing significant scattering or conversion, and the reduced air density gives **electrons** far fewer opportunities to collide. As a result, electrons generate much less bremsstrahlung before leaving the field region, further suppressing the upward γ-ray yield. Consequently, though electrons and positrons can run away easily at 9.3 km, they produce far fewer bremsstrahlung photons than in the denser layer at 7.3 km and leave the field region with minimal gamma-ray production.

Thus, even under fairly optimistic field strengths and particle injection rates, the absolute upward lepton and gamma–ray fluxes shown in Fig. 9 and Table 2 remain much lower than those measured during the ALOFT flights. Therefore, although RREA and positron deceleration effectively harden the upward spectrum, ambient cosmic rays alone are insufficient to provide a seed population for sustained RREA in the upper dipole.

## 6. The scheme of the feedback mechanism and estimation of infinite feedback possibilities.

In Dwyer's infinite feedback model, the mutual coupling between upward-moving electron avalanches and downward-moving positron avalanches creates a closed loop: each generation of gamma rays produces pairs that supply new seeds of opposite charge traveling in the direction that sustains the field-driven cascade. In theory, this process could lead to exponential growth in relativistic particles if the feedback coefficient (Γ) is equal to or exceeds 1. However, as shown in our simulations, most gamma rays are absorbed within the 2 km upper dipole (7.3–9.3 km), and the likelihood of both pair production and favorable emission direction is very low (<0.1%). As a result, the feedback coefficient remains several orders of magnitude below 1, preventing the development of a self-sustaining or "infinite" feedback loop.

Within a distance of L = 2 km, the air density varies significantly. Therefore, we divide the upper-dipole AEF into 10 slabs, each of 200 m, and estimate particle propagation step by step. The backward-going gamma-ray flux that reaches 9.3 km from the slab i can be written as:

$$\boldsymbol{N_{\gamma}^{\uparrow}(i \rightarrow 9.3\ km, from\ e^{+}\ and\ e^{-}\ RREA) = \{M_{e^{+}}^{\downarrow}(i, E_z, \gamma > 1.022\ MeV) \times P_{\mathrm{pair}}(h_i) \times f_{\uparrow}(e^{-}) \times T_i\,(9.3\ km)\} + \{M_{e^{-}}^{\uparrow}(i, E_z, \gamma > 1.022\ MeV) \times P_{\mathrm{pair}}(h_i) \times f_{\uparrow}(e^{-}) \times T_i\,(9.3\ \mathrm{km})\}} \quad \textbf{(1)}$$

Here $\boldsymbol{M_{e^{+}}^{\downarrow}(i, E_z, \gamma > 1.022\ MeV)}$ and $\boldsymbol{M_{e^{-}}^{\uparrow}(i, E_z, \gamma > 1.022\ MeV)}$ are the cumulative multiplication up to slab i (for positrons) and from slab i to 9.3 km (for electrons), i.e., the expected numbers of photons with energies above the pair–production threshold emitted by each avalanche. $P_{pair}\,(h_i)$is the probability that a photon converts into an e+e– pair at altitude h(i), $\boldsymbol{f_{\uparrow}(e^{-})}$ is the fraction of produced electrons that are emitted upward into the upper hemisphere, and $T_i$ (9.3 km) is the survival probability that these photons or leptons reach 9.3 km without being absorbed. Thus, Eq. (1) counts the number of backward–going gamma rays from slab i that are capable of producing new upward–moving electrons.

The infinite feedback (recursive) term in slab i can be written as

$$\boldsymbol{New\ cycle\ multiplication = N_{\gamma}^{\uparrow}(i \rightarrow 9.3\ km, from{:}\ e^{+}\ and\ e^{-}\ RREA) \times \{M_{e^{+}}^{\downarrow}(\mathrm{i}, \gamma, E_z) \times P_{\mathrm{pair}}(h_i) \times f_{\uparrow}(e^{-}) \times T_i\,(9.3\ \mathrm{km})} \quad \textbf{(2)}$$

Thus, the local feedback coefficient in slab i is

$$\boldsymbol{\Gamma_i = N_{\gamma}^{\uparrow}\ (i \rightarrow 9.3\ km) \times M_{e^{+}}^{\downarrow}(\mathrm{i}, \gamma, E_z) \times P_{\mathrm{pair}}(h_i) \times f_{\uparrow}(e^{-}) \times T_i\,(9.3\ \mathrm{km})} \quad \textbf{(3)}$$

In all our cases, $\Gamma_i \ll 1$, and the product over all slabs remains several orders of magnitude below unity.

The downward-moving positron avalanche produces gamma rays in each slab, and gamma-ray energy must exceed 1.022 MeV to enable pair production. The electron energy should be above 1 MeV to allow RREA to ascend in the opposite direction of the positron avalanche (a very small probability, but not zero). Downward-propagating positron RREA and upward-propagating electron RREA generate gamma rays that can produce pairs in each slab. Since this occurs in every slab, the contributions from all slabs must be summed. The multiplication in RREAs was derived from the e-folding length λ:

$$\boxed{\mathbf{M}_{e^{+}}^{\downarrow}(\boldsymbol{i, \gamma, E_z}) = \boldsymbol{e^{\frac{L(i)}{\lambda}}}} \quad \textbf{(4)}$$

Where the avalanche distance L(i) is 9.3 km − h(i) and h(i) is the height of the $i_{th}$ slab (middle of it). λ is the avalanche e-folding length at the given atmospheric electric field. For the upgoing electron avalanche the avalanche distance will be the same L(i) = 9.3 km – h(i):

$$\boxed{\mathbf{M}_{e^{-}}^{\uparrow}(\boldsymbol{i, \gamma, E_z}) = \boldsymbol{e^{\frac{\mathrm{L(i)}}{\lambda}}}} \quad \textbf{(5)}$$

The probability that at least one pair production event occurs within a slab of thickness 200 m is

$$\boldsymbol{P}_{\mathbf{pair}}(\boldsymbol{i}) = \mathbf{1} - \boldsymbol{e}^{-\frac{\mathbf{200}\,\boldsymbol{m}}{\mathbf{\Lambda pair(i)}}},\ \textbf{(6)}$$

where $\Lambda_{pair}$ (i) is the pair-production mean free path for photons with Eγ > 1 MeV expressed in meters. For 200 m ≪ Λpair(i), Eq. (6) reduces to the thin-slab approximation (Mayerhöfer et al., 2020) and can be written as

$$\boldsymbol{P_{pair}}(\mathbf{i}) \ \approx\ \mathbf{200m}/\boldsymbol{\Lambda_{pair}}(\boldsymbol{h_i}).\ \textbf{(7)}$$

The survival probability for photons of energy $E_\gamma$ to reach 9.3 km from slab i is given by

$$\boxed{\boldsymbol{T_i}(\mathbf{9.3}) = \boldsymbol{e}^{-\frac{\mathbf{L(i)}}{\boldsymbol{\Lambda(E_\gamma)}}}} \quad \textbf{(8)}$$

Here, $\Lambda(E_\gamma)$ is the effective attenuation length (m) for photons of energy $E_\gamma$ in air.

The Compton scattering cross-section and energy–angle redistribution are described by the Klein–Nishina formula (KN, Klein & Nishina, 1929), which strongly favors forward scattering at MeV energies. The angular distribution of leptons from $\gamma \rightarrow e^+e^-$ pair production follows the Bethe–Heitler theory (BH, Bethe & Heitler, 1934) and is strongly forward-peaked with a characteristic angle:

$$\boldsymbol{\theta_0} \ \simeq\ \frac{\mathbf{1}}{\boldsymbol{\gamma_e}} \ \approx\ \frac{\boldsymbol{m_e c^2}}{\boldsymbol{E_e}}, \quad \textbf{(9)}$$

A conservative upper-bound for the backward-hemisphere fraction can be approximated as

$$\boldsymbol{f}_{\uparrow}(\,\boldsymbol{e} \mid \boldsymbol{E_e}\,) \ \approx\ \boldsymbol{e}^{-\frac{(\boldsymbol{\pi}/\mathbf{2})^2}{\mathbf{2}\,\boldsymbol{\theta_0^2}}} = \boldsymbol{e}^{-\frac{\boldsymbol{\pi^2}}{\mathbf{8}\boldsymbol{\gamma_e^2}}}\,, \qquad \textbf{(10)}$$

We combine the spectral weighting and backward-hemisphere acceptance into a single small factor f↑↓, representing the probability that a lepton emerges opposite to the photon direction (for positron RREA upward in 7.3-9.3 km and for electron RREA downward from 9.3 km).

Using an optimistic pair-production probability $P_{pair}(i) \approx 10^{-2}$ per 200 m slab (NIST XCOM database, Berger et al., 2010, Motz et al., 1969), a backward-hemisphere probability f↑↓ ≈$10^{-5}$, consistent with angular distributions from Haug (1975), a gamma-ray attenuation factor T=0.15, and ten slabs along the 2-km AEF, we obtain a yield of upgoing gamma rays per incident positron or electron:

$$\mathbf{P}\big(\mathbf{9.3},\ \boldsymbol{E_\gamma} > \mathbf{1.022}\ \boldsymbol{MeV}\big) \ =\ \mathbf{10} \times \boldsymbol{P}_{\mathbf{pair}} \times \boldsymbol{f}_{\uparrow\downarrow} \times \boldsymbol{T}_{\mathbf{1.022}\ \boldsymbol{MeV}} \ = \mathbf{1.5}\ . \times \mathbf{10^{-7}} \quad \textbf{(11)}$$

For a rough estimate of the gamma-ray output, we also adopt an optimistic RREA multiplication factor of $10^4$ for both positron- and electron-driven avalanches. Under these assumptions, the total number of upward gamma rays per initial seed lepton is

$$\boldsymbol{N_{\gamma}^{\uparrow}(E_z, 9.3\ km, from\ e^{+}\ and\ e^{-}\ RREA) = 2 \times \{M_{e^{\uparrow\downarrow}}^{\downarrow}(E_z, \gamma > 1.022\ MeV)\ \times\ P(9.3)\} = 3\ \times 10^{-4}} \quad \textbf{(12)}$$

Each surviving upgoing gamma ray can again generate an electron-positron pair with a redirected downward positron, with probability $P_{pair}(9.3)$ * $f^{\downarrow}$ $(e^{+})$ . Thus, the total expected number of recycling positrons is:

$$\boldsymbol{N_{e^{+}}^{\downarrow}(recycling) =\ N_{\gamma}^{\uparrow}(around\ 9.3\ km, from\ e^{+}\ and\ e^{-}\ RREA) \times P_{\mathbf{pair}}(9.3)\ \times\ f^{\downarrow} \approx 10^{-10}} \quad \textbf{(13)}$$

This is the total feedback-return $\Gamma_{total}$ per cycle (feedback coefficient). Even under deliberately optimistic assumptions, the feedback coefficient remains of order $10^{-10}$, demonstrating that self-sustaining relativistic feedback discharge is impossible in the considered configuration.

During CORSIKA simulations of the upper dipole at altitudes of 7.3–9.3 km, we examined all physically plausible scenarios for upgoing particle seeds within the RFD framework proposed by Dwyer (2003, 2025). Our approach provides a quantitative framework for interpreting experimental TGE spectra, modeling particle transport in AEF, and linking lepton interactions to Monte-Carlo simulations.

A key difference between the original RFD formulation and our treatment lies in the explicit evaluation of the full return-probability chain for feedback particles. In the RFD model, the feedback factor is defined as the product of avalanche multiplication and an effective return efficiency. In contrast, the detailed probabilities governing gamma-ray survival, pair production, and emission direction are not evaluated explicitly. In reality, a γ ray with energy above 1.022 MeV reaching 9.3 km must both survive propagation through several hundred meters of atmosphere and undergo pair production with the positron emitted downward into the high-field region. Because pair production at MeV energies is strongly forward-peaked, the probability of downward positron emission is extremely small ($\boldsymbol{f^{\downarrow}(e^{+})} < 10^{-5}$), $E_{\gamma} > 1$ MeV. When combined with the pair-production probability per 200-m slab ($P_{pair} \approx 10^{-2}$), the total recycling probability becomes $\Gamma \approx 10^{-10}$, even under optimistic assumptions. Consequently, the effective feedback coefficient remains many orders of magnitude below unity, and no self-sustaining feedback loop is possible. The absence of any cycling behavior in multiple independent CORSIKA simulations supports this conclusion.

These results are consistent with earlier numerical studies (Stadnichuk et al., 2017; Zelenyi et al., 2022) and with Aragats measurements (Chilingarian, 2017b; Chilingarian et al., 2024a), which show random arrival times of avalanche particles rather than the continuous flux expected from an active RFD.

Upgoing $E_\gamma$>50 keV photons produced from "inner seeds" (ambient CR γ and positrons) are influenced by transport, not directly by the electric field. The key point is that 50 keV photons have a short attenuation length in air, $\lambda_{50keV} \approx (1–1.3)\times10^2$ m, so essentially only about 200 meters below 9.3 km can contribute to the escaping flux. Therefore, the top roughly 200 meters acts as an "albedo/converter layer": any photon field (ambient or RREA-generated) entering this layer can produce escaping >50 keV photons through a final Compton scatter (and a small backward-bremsstrahlung contribution), whereas photons produced deeper cannot escape at 50 keV; this provides a conservative baseline. In the 2 km case, the down-going positrons trigger a positron RREA in a super-threshold field ($E_z$=2.1 kV/cm), producing an enhanced MeV γ field throughout the column. MeV photons can be Compton-scattered upward with far less attenuation than 50 keV photons, generating additional upgoing >50 keV photons.

A physically justified factorization of Comptonization for the 200 m slab is

$$\boldsymbol{N^{\uparrow}_{\gamma>50\,keV}(9.3\ km, from\ e^{+} and\ \ \gamma\ CR\ fluxes) = P_{\mathbf{Compt}} \times\ f_{\uparrow>50keV}\ \times T_{50} + M^{\downarrow}_{e^{+}}(E_z, \gamma > 50\ keV)\ \times\ P^{\uparrow}_{RREA}(9.3,\ E_{\gamma>50\ keV})} \quad \textbf{(14)}$$

Where $P_{Compt}$ is the spectrum-averaged probability that an incident photon undergoes at least one Compton interaction in a 200 m slab. A plausible range for $P_{Compt}$ is approximately 0.3 to 0.8. The parameter $f_{\uparrow>50keV}$ represents the fraction of gamma rays that are redirected upward and remain above 50 keV; a plausible range for $f_{\uparrow>50keV}$ is approximately 0.3 to 0.7. $T_{50keV}$ is approximately 0.20. The gamma yield per CR gamma ray is around 0.02 to 0.11, consistent with CORSIKA simulations (see Tab. 2). We assume that in the positron RREA starting at 9.3 km, the gamma-ray multiplication factor M equals 10,000, and that gamma rays with energy greater than 50 keV produced in these RREAs will interact in air and be redirected into the upper hemisphere with a probability $P_{RREA} \approx 3 \times 10^{-5}$. We adopt PRREA as an upper-bound estimate for the backscatter and survival probability of MeV photons produced in the positron-initiated avalanche to reach the 9.3 km boundary, considering the relatively long attenuation length of MeV photons at 7-9 km altitude and the significantly large-angle Compton component observed in CORSIKA simulations. Therefore, we assign this probability three times higher than that of redirected gamma rays with energy above 1.022 MeV. Adding these factors together, we estimate a value of 0.4 per seed positron, which aligns with CORSIKA simulation results.

The DSM, in which backscattered gamma rays from electron RREA in the lower dipole initiate RREA in the upper dipole, provides many more seed particles for further multiplication in the upper dipole. Because electrons produced by γ interactions propagate in the same direction as the incident photons, no angular penalty applies, enabling efficient multiplication. For an ambient electron flux of $10^3$ m$^{-2}$ s$^{-1}$ and an RREA multiplication factor of 4000, the resulting 50-keV gamma-ray flux escaping the upper dipole reaches approximately $1.8 \times 10^6$ m$^{-2}$ s$^{-1}$, surpassing the contribution from inner ambient seeds by roughly three orders of magnitude, see Table 3, where we present the CORSIKA simulated fluxes and yields along with theoretical estimates.

**Table 3. Summary of quantitative results on the considered mechanisms of the gamma-ray flux from the upper dipole**

| Mechanism | Flux ($m^{-2}\ s^{-1}$) or yield per seed | Notes |
|---|---|---|
| Ambient positron. Gamma ray, and electron fluxes in upper and lower dipoles | $1 \times 10^3$ | Approximate estimate from EXPACS (Sato, 2016) |
| CORSIKA yield of > 1 MeV gamma-ray from the lower dipole | 0.45 | Per seed electron |
| RREA multiplication factor for a > 1 MeV gamma-ray entering 7.3 km from the lower dipole | 4000 | Gamma ray multiplication in the electron RREA in upper dipole |
| Flux of 50 keV gamma-ray to open space by DSM | $1.8 \times 10^6$ | DSM flux estimate using CORSIKA-derived yield/multiplication inputs: 1000 × 0.45 × 4000 |
| Corsika yield for 50 keV gamma-ray per ambient positron at 9.3 km | 0.375 | |
| Corsika yield for 50 keV gamma-ray per ambient gamma-ray at 9.3 km | 0.094 | |
| Theoretical estimate of 50 keV gamma-ray yield per inner ambient seed | 0.475 | Per seed positron and gamma ray under optimistic assumptions |
| CORSIKA flux for 50 keV gamma-ray to open space from inner seeds | $4.75 \times 10^2$ | Estimated flux from inner seeds: 1000 x 0.475 |
| RFD infinite-feedback | $\approx 10^{-10}$ | Recycling probability negligible; infinite feedback impossible |

The >50 keV DSM flux can be compared with measurements from the ALOFT mission. The spectrometer onboard ALOFT has a detection threshold of about 300 keV (Ostgaard et al., 2024). Therefore, we adjust the DSM flux to match the ALOFT spectrometer's threshold, resulting in flux ≈ (1.1–1.3) × $10^6\ m^{-2}\ s^{-1}$. This adjustment ensures that the DSM estimates in the 50 keV–MeV range fall within the BGO sensitivity band (Marisaldi et al., 2024). The 10 km flux can also be compared to the 20 km flux because we position the tripole at 5.3–9.3 km, in line with the Aragats location, and during tropical storms, the storm height can be significantly higher. The corrected value aligns with the bright gamma-ray glows observed by ALOFT and falls between the lower and upper limits of ALOFT gamma fluxes, as shown in Table 4.

**Table 4. The comparison of DSM flux estimates and registered ALOFT fluxes**

| Source / Event | Flux at 10 km (≥300 keV) ($m^{-2}$ $s^{-1}$) | BGO Count Rate (≥300 keV) ($m^{-2}$ $s^{-1}$) | iSTORM Count Rate (≥300 keV) ($m^{-2}$ $s^{-1}$) |
|---|---|---|---|
| DSM Upper-Dipole (simulation, 10 km) | 1.1–1.3 × $10^6$ | | |
| ALOFT (low) | n/a | ≈ 2.3 × $10^2$ | ≈ 8.1 × $10^1$ |
| ALOFT (high) | n/a | ≈ 7.8 × $10^5$ | ≈ 2.7 × $10^5$ |

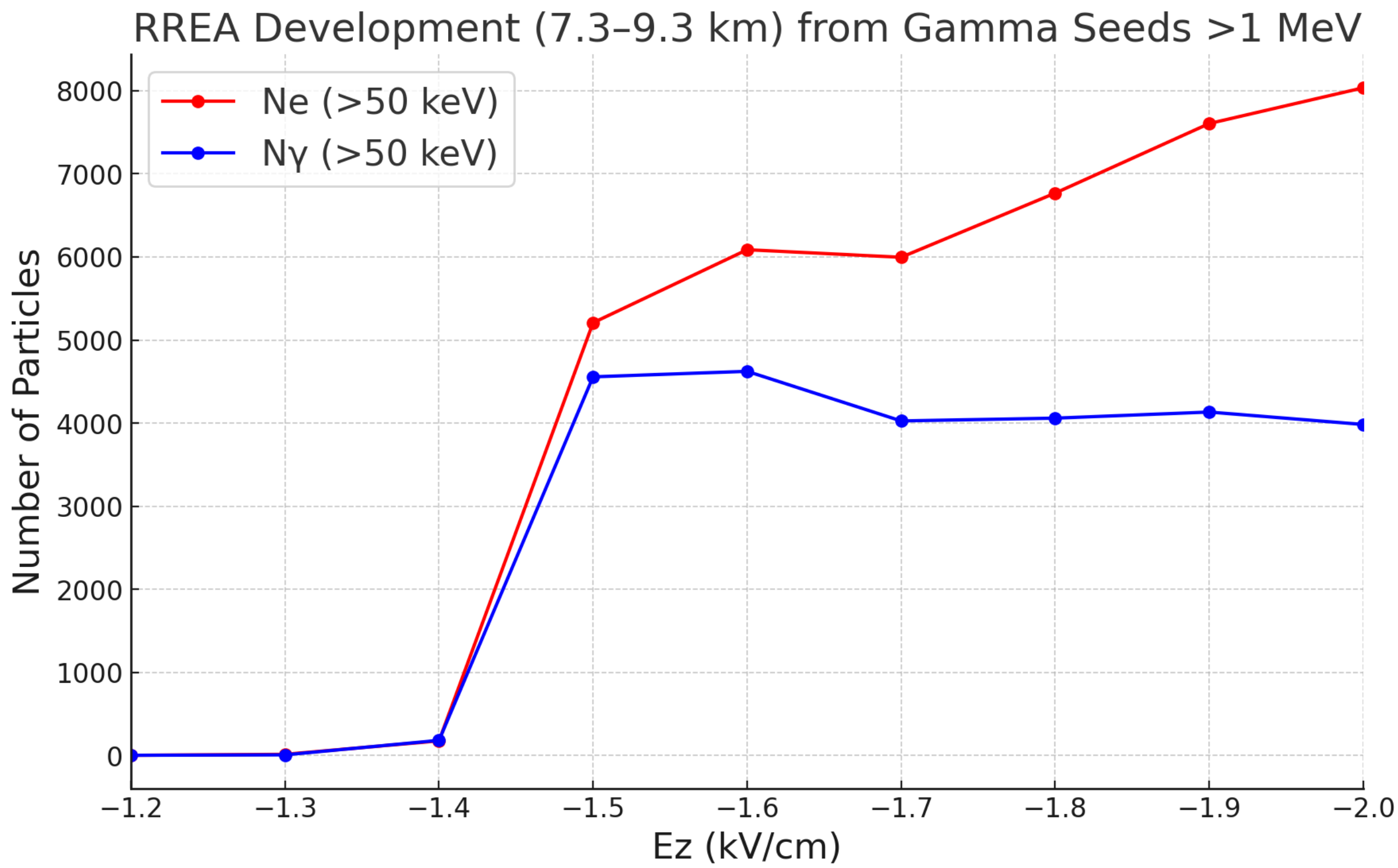

**Figure 10. Multiplication of RREA particles per seed gamma ray with energy larger than 1 MeV in the upper dipole.**

Another interesting result shown in Fig. 10 is that, as the AEF increases, the gamma rays produced in electron RREA reach a peak at 1.5 kV/cm and then level off, even decreasing with further increases of the electric field. For seed energy $E\gamma > 1$ MeV injected at the lower boundary of the upper dipole, the number of secondary gamma rays produced by the avalanche increases with field strength up to an optimum (~1.5 kV/cm) and then saturates or decreases. This non-monotonic behavior is expected in rarefied air: while stronger fields increase the characteristic energies of runaway leptons, the interaction probability per unit path length (bremsstrahlung and pair production) scales with air density. At sufficiently high E, more particles traverse the region with fewer interactions, so the gamma-ray yield per unit length does not continue to grow. This motivates using altitude-dependent 'optimal' fields in simulations for photon production rather than adopting unnecessarily high Ez values.

Figure 11 summarizes the altitude dependence of the electric field that maximizes the gamma-ray yield in the upper dipole. The decreasing optimal Ez with altitude reflects the same density control seen in Figure 10. At higher altitudes, a smaller field is sufficient to reach energies where further field increases no longer translate into more photon production because the interaction rate becomes density-limited. In practice, this result supports selecting, in simulations, Ez close to the altitude-dependent optimum when estimating observable gamma-ray fluxes, rather than maximizing Ez/Eth alone.

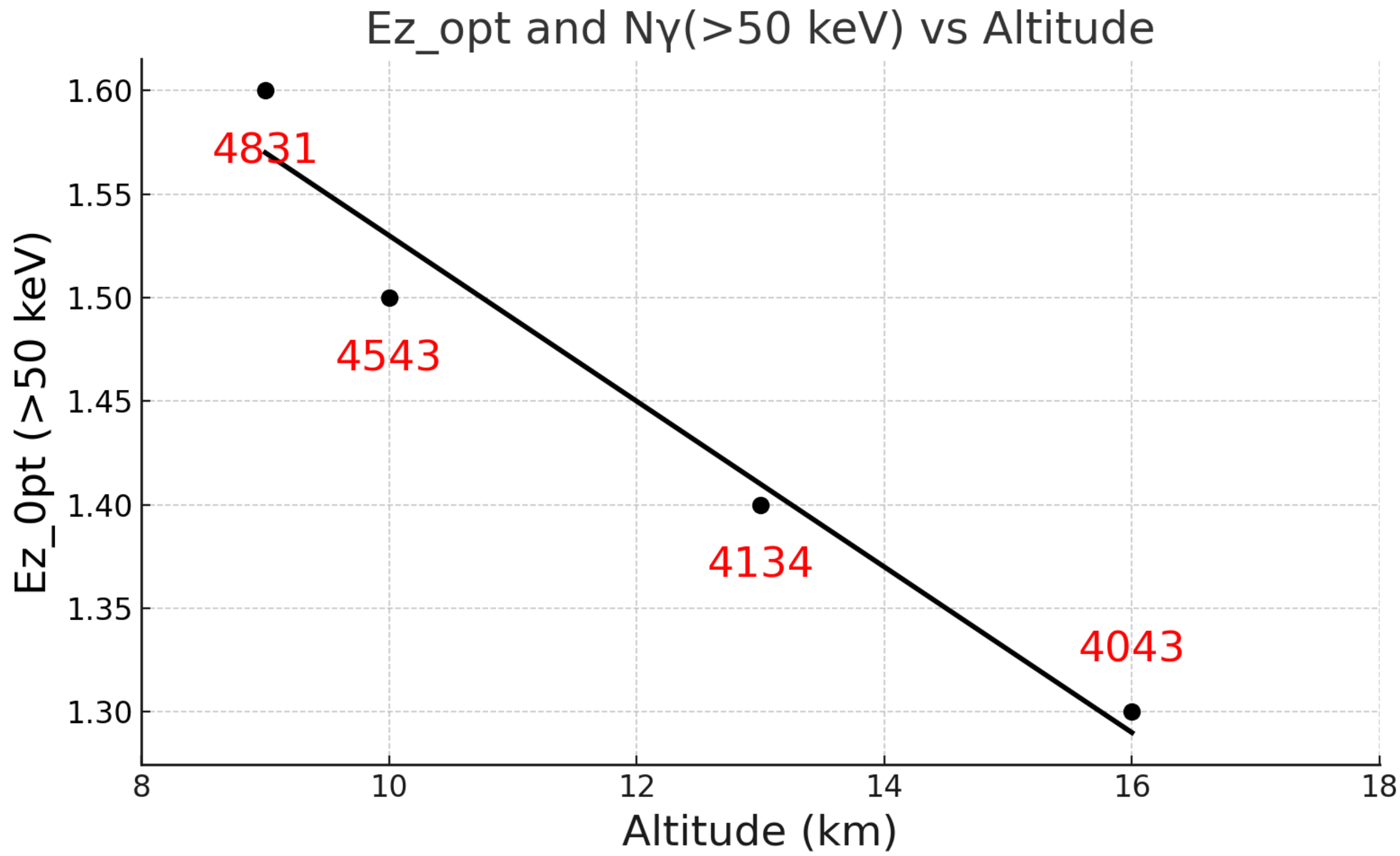

**Figure 11. Altitude dependence of optimal AEF (for maximum gamma rays in the avalanche). Red numbers denote achieved maxima of gamma-ray flux among the tested 1-2 kV/cm electric fields.**

Therefore, when directional constraints, survival probabilities, and realistic interaction physics are taken into account, a self-sustaining relativistic feedback discharge is not achievable. The dual-stage seeding mechanism provides a physically consistent and quantitatively supported alternative explanation for intense high-altitude gamma-ray emissions.

## 7. Model of the DTGF origination

Wada et al. (2025) propose a DTGF model based on converging positive and negative leaders, which was observed on January 30, 2023, in Kanazawa, Japan. A 31-μs gamma-ray burst detected by scintillators was accompanied by lightning leader detection, resulting in cloud-to-ground lightning (-CG). A downward negative leader initiated at an altitude of 2 to 3 km extends toward the ground. The upward positive leader propagated from the tower, see Fig. 12). The first gamma rays of the downward TGF started 31 μs before the return stroke and reached saturation level 17 μs before RS. This suggests that the TGF occurred when the upward and downward leaders were approaching and just before they collided.

The authors assume that a strong electric field in the narrow gap between approaching leaders initiates RREA. However, they overlook fundamental constraints imposed by avalanche physics on the spatial extent of the electric field responsible for RREA. The observed DTGF duration, $\Delta t \approx 31$ μs, along with the stepped leader approach speed ($v \approx 1.8 \times 10^6$ m/s, Wada, 2025), suggests that electron acceleration continues when the gap distance is less than: $\Delta h = v \cdot \Delta t \approx 1.8 \cdot 10^6$ m/s $\cdot\, 31 \cdot 10^{-6}$ s $\approx 56$ m.

The characteristic avalanche (e-folding) length for RREA at near-ground pressure is approximately 70 meters (Celestin & Pasko, 2011; Dwyer et al., 2012); also see discussion at the beginning of section 5. Efficient avalanche multiplication, necessary for observable DTGF production, requires several e-folding lengths to be available before the rapidly closing gap between the two leaders. This implies that if the acceleration region were confined to the leader–leader gap, RREA and significant gamma-ray emission would not be possible.

The DTGF terminates when the potential difference is neutralized by leader reconnection. In this model, the upward positive leader does not initiate the DTGF; instead, it terminates it by neutralizing the potential difference between the negative leader and Earth, explaining the abrupt cessation of the gamma-ray burst. The acceleration process is thus sustained in the full vertical leader–ground gap, rather than in the narrow region between the approaching leader tips (see Figure 9).

Multiyear DTGF observations in Uchinada, Japan, are also linked to negative cloud-to-ground lightning, demonstrating that gamma-ray fluxes span a vast area of at least 100 km² around the lightning flash (see Fig. 1 of Ortberg et al., 2024). This raises a fundamental question: how can a small negative leader tip, only a few cm² in size, influence atmospheric electric fields over 5 kilometers to initiate RREA across extensive cloud volumes (see Fig.12)?

In contrast, the downward propagation of a negative leader in a pre-critical dipole field between the main negative (MN) layer and image charges near the ground can induce a non-local reorganization of the AEF. Thus, the negative leader acts not as a direct accelerator but as a trigger for large-scale field transformation, displacing equipotential surfaces and enhancing preexisting subcritical fields over vast volumes. As the negative leader descends, it effectively 'pulls down' the potential of the MN layer, compressing the equipotential surfaces below it. This deformation increases the local electric field between the MN layer and the ground, transforming previously subcritical regions into zones that extend 800 m vertically and 10 km horizontally, capable of initiating RREA. This mechanism, demonstrated by the detection of hundreds of TGEs worldwide (Chilingarian et al., 2025), allows extensive areas, spanning many square kilometers, to experience a deepening of negative AEF from modest to supercritical levels. Thus, the DTGFs observed in Uchinada receive a natural explanation within the large-scale extension of AEF in the lower dipole, rather than assuming an enigmatic source of $10^{19}$ electrons above 1.5 km from the surface.

Recent observations by the Telescope Array Surface Detector (TASD) also linked DTGFs to downward negative leaders traveling at a velocity greater than 3 × 10^6 m/s, which result in -CG-lightning flashes (Kieu et al., 2025). They didn't report upward lightning leaders but confirmed the relationship between TGF production and fast downward negative leaders, as suggested by Wu et al. (2021).

TGEs are initiated in the strong AEFs, surpassing the RREA threshold values, and their development—sometimes lasting for a few tens of minutes—was not accompanied by lightning activity. However, the inverted intracloud (-IC), -CG, or hybrid flashes (beginning as -IC and then turning into -CG) often ended the TGEs. Recent analysis of 163 TGEs terminated by lightning flashes at Aragats demonstrates that the share of -CGs, -ICs, and hybrid flashes is significantly higher, at 92.7%, compared to the share of normal polarity ICs, which is only 11.3% (Chilingarian, Rakov, and Khanikyannc, 2024).

Thus, the proposed DTGF model aligns with DTGF observations at Uchinada, Utah, and recent gamma glows (Marisaldi et al. 2024; Ostgaard et al., 2024), and multi-year observations of thunderstorm ground enhancements (TGEs) by the SEVAN network. (Chilingarian et al., 2010; 2011). These observations reveal that high-energy particle bursts can affect regions covering many square kilometers before lightning occurs.

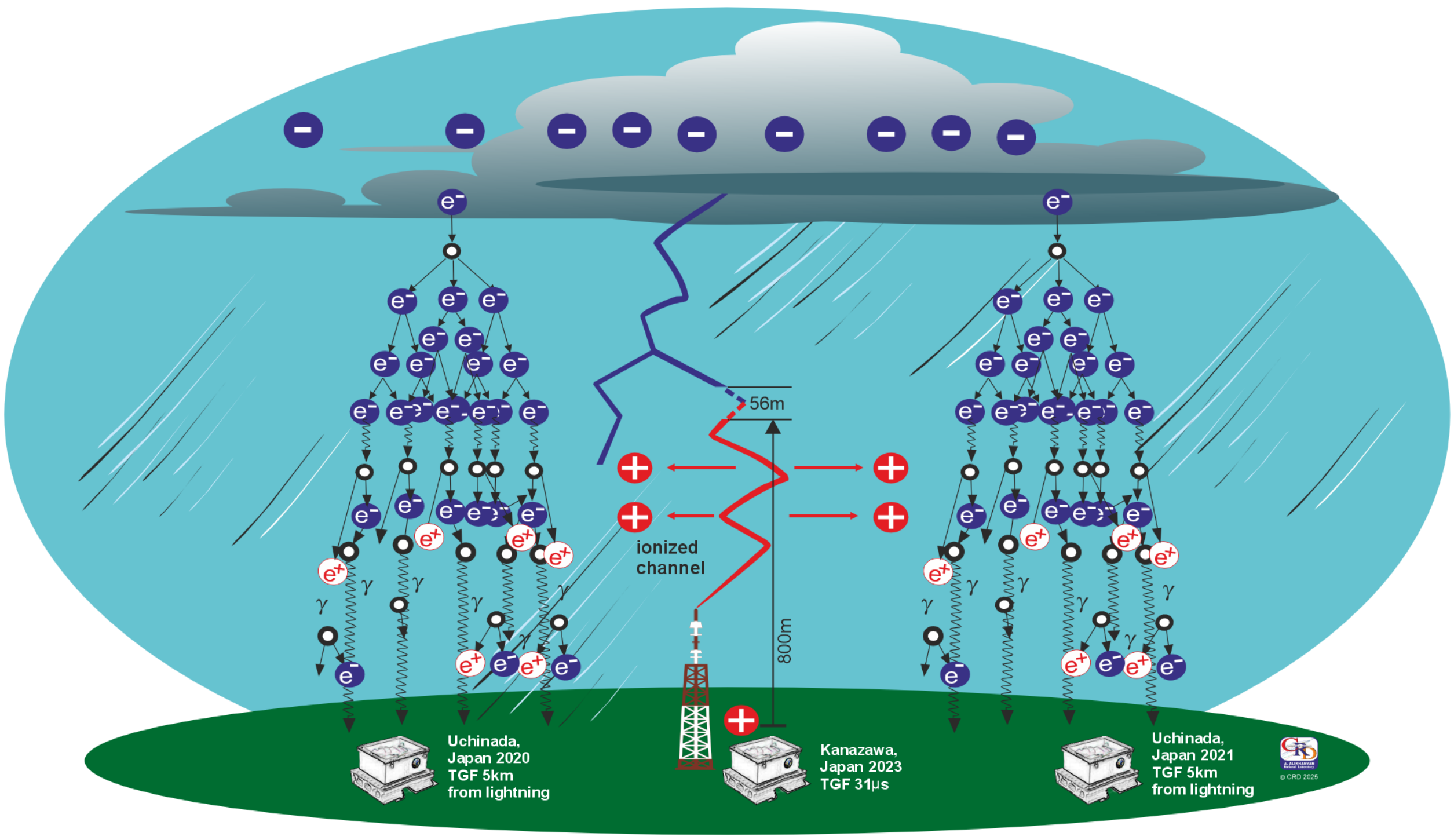

**Figure 12. Revised DTGF model. Approaching lightning leaders (center) induce non-local reorganization of the atmospheric electric field (AEF), transforming previously subcritical regions into RREA-capable zones. RREA regions are shown at distances of ~5 km from the lightning channel, consistent with observations in Uchinada (2020–2021).**
**Blue – downward negative leader; red – upward positive leader.**

## 7. Discussion and Conclusions

Relativistic runaway electron avalanches cause thunderstorm ground enhancements, gamma-ray glows, and both upward and downward terrestrial gamma flashes. Recognizing the common origin for microsecond TGFs, minute-long gamma glows, and TGEs indicates progress toward accepting RREA and EAS as universal physical processes that produce increased particle fluxes in both the lower and upper atmosphere. For this ambitious program, the most challenging task is identifying the seed sources for the upward gamma rays from the upper dipole.

Several mechanisms have been proposed for the origin of seed electrons. The photoelectric model (Pasko et al.,2025) suggests that X-ray–driven photoionization produced within the relativistic feedback framework photo-ionize air molecules, liberating electrons even below the runaway threshold. While effective in explaining lightning initiation at leader tips, this model operates on very short timescales (μs–ms) and does not directly account for sustained, seconds-minutes-long MeV gamma glows observed far from leader channels.

The cold-runaway model (Celestin & Pasko, 2011) mainly examines runaway electron production at lightning-leader tips, where very intense, localized electric fields can accelerate low-energy ("cold") electrons to relativistic energies. Their model aims to explain the high initial electron fluxes that seed leader-associated TGFs and naturally functions on microsecond timescales within tens of meters around the leader channel.

Dwyer's relativistic feedback model (RFDM, Dwyer, 2003) is a single-region, closed-loop mechanism (infinite feedback) where the ambient population of cosmic rays generates backward-moving positrons that re-enter the same high-field layer, thereby sustaining the avalanche.

In contrast, the proposed double-stage model involves two regions and a feedforward process: the lower dipole supplies MeV photon seeds to the upper dipole, which are converted into upward electrons that trigger RREA in the upper dipole. This setup naturally accommodates kilometer-scale field gaps, large horizontal footprints, and lightning-silent, minutes-long glows observed by aircraft and balloons. Relativistic feedback may still operate within each dipole and slightly modify the local multiplication, but it will not close the loop across the gap and does not alone explain the altitude, duration, and areal extent of the upward glows. The altitude range, intensity, duration, and frequent radio silence reported in the ALOFT campaign align with the predicted upward transport and amplification in the upper dipole described by the dual-stage model.

Observed DTGFs cover large areas and last tens of microseconds, with duration, intensity, and spatial scales that are hard to reconcile with RREA confined to a narrow leader–leader gap. The e-folding length near ground pressure suggests that several e-foldings are needed. However, a 56 m length that closes in 31 μs cannot provide enough path for avalanche development and strong gamma multiplication. Instead, in our view, a descending negative leader in a pre-critical lower-dipole field reorganizes the AEF non-locally, deepening subcritical regions into supercritical ones over hundreds of meters vertically and kilometers horizontally. This process enables RREA over a wide footprint. Reconnection then quickly quenches the potential drop, explaining the sudden end of DTGFs. This non-local field reorganization naturally links DTGFs with TGEs and glows, all of which are manifestations of RREA in different geometries.

Nevertheless, we fully acknowledge that *special* meteorological conditions, such as Japanese winter thunderstorms, may support a different class of gamma-ray events. Although downward TGFs associated with compact lightning strokes have recently been reported over Japan (Wu et al., 2025), long-term measurements at Aragats and the combined optical-particle observations from ASIM and ALOFT show that TGEs and gamma-ray glows are not triggered by lightning; lightning activity typically suppresses or terminates them. Yet, no confirmed mechanism exists by which lightning processes could provide the required upward-directed MeV seed electrons for RREA.

Various physical processes drive atmospheric particle fluxes, making it essential to accurately identify each one. Since TGE research began at Aragats in 2010, nearly 1,000 TGEs have been recorded on mountaintops across Eastern Europe, Japan, Russia, Germany, and Armenia (Chilingarian et al., 2025). Early observations of gamma-ray bursts are already reviewed and discussed comprehensively in Chilingarian (2024); here, we focus on the particle-flux characteristics of gamma glows in lower and upper dipoles as revealed by recent ALOFT and ASIM measurements. Recent observations have been made at Mt. Hermon in Israel (Mauda et al., 2025) and in Finland (Kärkkäinen et al., 2024). Measurements of TGE electrons and gamma rays confirmed RREA as a viable electron accelerator covering many square kilometers on Earth's surface. Along with simulations, these findings have provided a detailed understanding of RREA development and cloud charge structures in the lower atmosphere. According to DSM, the seed electrons for TGE are free electrons from EASs, while gamma glow seeds originate from RREA in the lower atmosphere in the form of reverting gamma rays. The atmospheric electric field in the upper and lower dipoles, ranging from 1.4 to 2.1 kV/cm, is sufficient to trigger RREAs at altitudes of 3-15 km. Enhanced electric fields and seed particles moving both downward and upward can explain the energy spectra of electrons and gamma rays in measured TGEs, as well as the intensity of gamma glows. During DTGF, the rearrangement of the electric field through lightning leader interactions increases pre-critical AEF to levels sufficient for the runaway process.

Gaining a deeper understanding of different charge structures and the hierarchy of particle acceleration mechanisms will clarify the high-energy phenomena observed in thunderstorms and their implications for atmospheric science.

**Notations:**

The following terminology pertains to the atmospheric fluxes of elementary particles produced during thunderstorms:

- Terrestrial gamma-ray flashes (TGFs) are short bursts of gamma radiation detected by orbiting gamma-ray observatories. They last for tens of microseconds, originate from thunderstorms in equatorial regions, and are observed by the orbiting gamma-ray observatories positioned 400 to 700 kilometers above the source.

- TGEs (thunderstorm ground enhancements) refer to intense and prolonged particle fluxes observed on the Earth's surface, lasting from seconds up to several tens of minutes. These fluxes originate from the same RREA process; the accelerating electric fields are located just above the detectors (sometimes 25-100 m), allowing for a detailed study of the energy spectra of electrons and gamma rays, as well as the charge structures of thunderclouds.
- Gamma glows are bursts of gamma radiation from RREAs in the upper dipole detected high in the atmosphere by instruments carried on balloons or aircraft. These events can last from tens of seconds to several minutes and typically conclude with a lightning strike.
- The gamma-ray enhancements observed at the Earth's surface are sometimes also called gamma glows because thunderclouds, due to their high altitude, exclusively detect gamma rays and no other types of radiation particles (Wada et al., 2021).

- The lightning leaders can initiate downward TGFs by triggering a large-scale field transformation, displacing equipotential surfaces and enhancing preexisting subcritical electric fields over vast thunderstorm volumes.
- AEF – Atmospheric Electric filed
- Dual-stage model (DSM) in which RREA development in the lower dipole generates upward-directed gamma rays that seed RREA in the upper dipole.
- Relativistic Feedback Discharge (RFD) cycling infinite loop model in the upper dipole